\documentclass[preprint,journal]{vgtc}       

\ifpdf
  \pdfoutput=1\relax                   
  \usepackage{graphicx}                
  \DeclareGraphicsExtensions{.pdf,.png,.jpg,.jpeg} 
\else
  \usepackage{graphicx}                
  \DeclareGraphicsExtensions{.eps}     
\fi%

\graphicspath{{figures/}{pictures/}{images/}{./}} 

\usepackage{microtype}                 
\PassOptionsToPackage{warn}{textcomp}  
\usepackage{textcomp}                  
\usepackage{mathptmx}                  
\usepackage{times}                     
\usepackage{cite}                      
\usepackage{tabu}                      
\usepackage{ccicons}
\usepackage{booktabs}                  
\usepackage{wasysym}
\usepackage{fontawesome}
\usepackage{float}
\usepackage{dblfloatfix} 
\usepackage{lettrine}
\usepackage{hyperref}
\usepackage[normalem]{ulem}

\usepackage[dvipsnames]{xcolor}
\definecolor{design1}{RGB}{31,119,180}
\definecolor{design2}{RGB}{255,127,14}
\definecolor{design3}{RGB}{44,160,44}

\definecolor{removed}{RGB}{222,222,222}

\newcommand{\revision}[1]{\textcolor{design2}{#1}}
\renewcommand{\revision}[1]{#1}
\newcommand{\carmen}[1]{\textcolor{red}{(CH: #1)}}
\newcommand{\robert}[1]{\textcolor{purple}{(RK: #1)}}
\newcommand{\matt}[1]{\textcolor{blue}{(MB: #1)}}

\newcommand{\hli}[1]{\tiny{\colorbox{design1}{\textsf{\textcolor{white}{\textbf{#1}}}}}\normalsize}
\newcommand{\hlii}[1]{\tiny{\colorbox{design2}{\textsf{\textcolor{white}{\textbf{#1}}}}}\normalsize}

\newcommand{\caphli}[1]{\tiny{\colorbox{design1}{\textsf{\textcolor{white}{\textbf{#1}}}}}\scriptsize}
\newcommand{\caphlii}[1]{\tiny{\colorbox{design2}{\textsf{\textcolor{white}{\textbf{#1}}}}}\scriptsize}

\renewcommand{\carmen}[1]{}
\renewcommand{\robert}[1]{}
\renewcommand{\matt}[1]{}

\newcommand{\etal}{et al.}
\newcommand{\eg}{e.g.,\ }

\newcommand{\bstart}[1]{\vspace{1mm} \noindent{\textbf{#1.}}}
\newcommand{\istart}[1]{\vspace{1mm} \noindent{\textit{#1.}}}

\DeclareRobustCommand{\mHoundstooth}{%
  \begingroup\normalfont
  \includegraphics[height=\fontcharht\font`\B]{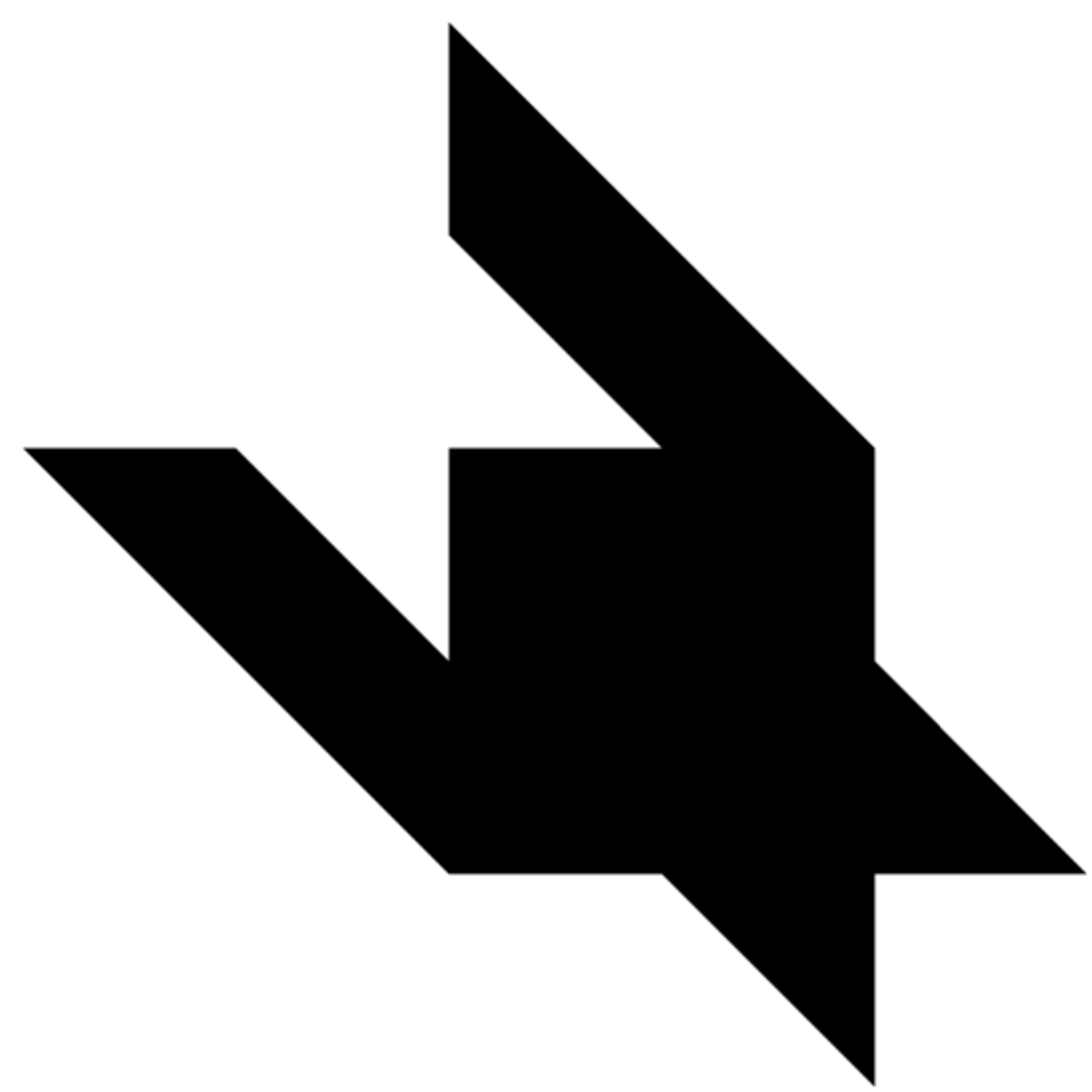}%
  \endgroup
}

\DeclareRobustCommand{\mDrop}{%
  \begingroup\normalfont
  \includegraphics[height=\fontcharht\font`\B]{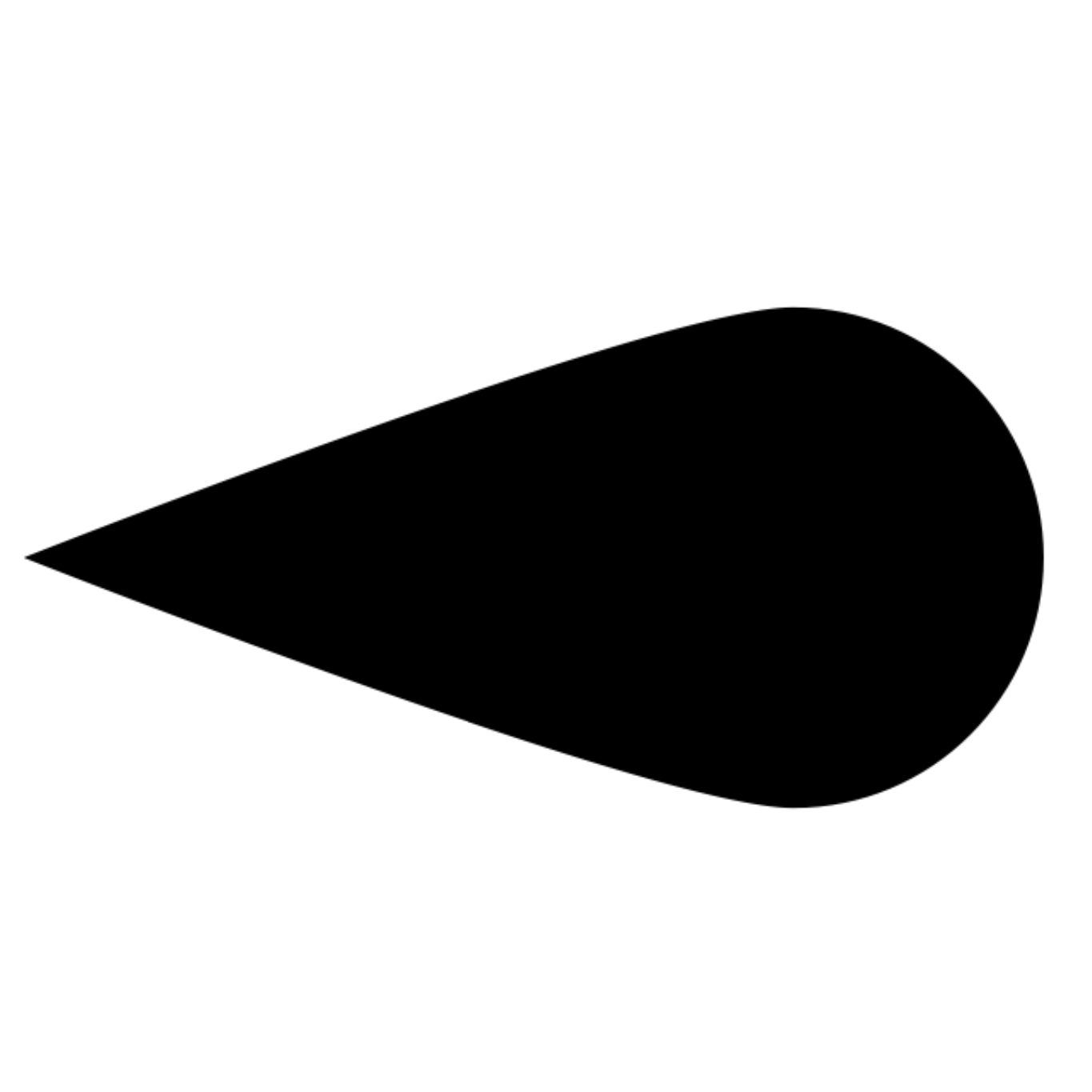}%
  \endgroup
}

\preprinttext{Authors' version of the paper accepted to IEEE VIS 2021; to appear in IEEE Transactions on Visualization \& Computer Graphics.}

\onlineid{1064}

\vgtccategory{Research}
\vgtcpapertype{Representations \& Interaction}

\title{Generative Design Inspiration for Glyphs with Diatoms}

\author{Matthew Brehmer, Robert Kosara, and Carmen Hull}
\authorfooter{
\item
 Matthew Brehmer and Robert Kosara are with Tableau Research. E-mail: \{mbrehmer,rkosara\}@tableau.com.
\item
 Carmen Hull is with the University of Calgary. E-mail: carmen.hull@ucalgary.ca.
}

\shortauthortitle{Brehmer \MakeLowercase{\textit{et al.}}: Generative Design Inspiration for Glyphs with Diatoms}

\abstract{We introduce Diatoms, a technique that generates design inspiration for glyphs by sampling from palettes of mark shapes, encoding channels, and glyph scaffold shapes.
Diatoms allows for a degree of randomness while respecting constraints imposed by columns in a data table: their data types and domains as well as semantic associations between columns as specified by the designer.
We pair this generative design process with two forms of interactive design externalization that enable comparison and critique of the design alternatives.
First, we incorporate a familiar \emph{small multiples} configuration in which every data point is drawn according to a single glyph design, coupled with the ability to page between alternative glyph designs. 
Second, we propose a \emph{small permutables} design gallery, in which a single data point is drawn according to each alternative glyph design, coupled with the ability to page between data points.
We demonstrate an implementation of our technique as an extension to Tableau \revision{featuring three example palettes}, and to better understand how Diatoms could fit into existing design workflows, we conducted interviews and chauffeured demos with 12 designers.
Finally, we reflect on our process and the designers' reactions, discussing the potential of our technique in the context of visualization authoring systems.
Ultimately, our approach to glyph design and comparison can kickstart and inspire visualization design, allowing for the serendipitous discovery of shape and channel combinations that would have otherwise been overlooked.
} 


\CCScatlist{ 
 \CCScat{K.6.1}{Management of Computing and Information Systems}%
{Project and People Management}{Life Cycle};
 \CCScat{K.7.m}{The Computing Profession}{Miscellaneous}{Ethics}
}

\teaser{
  \centering
  \includegraphics[width=\linewidth]{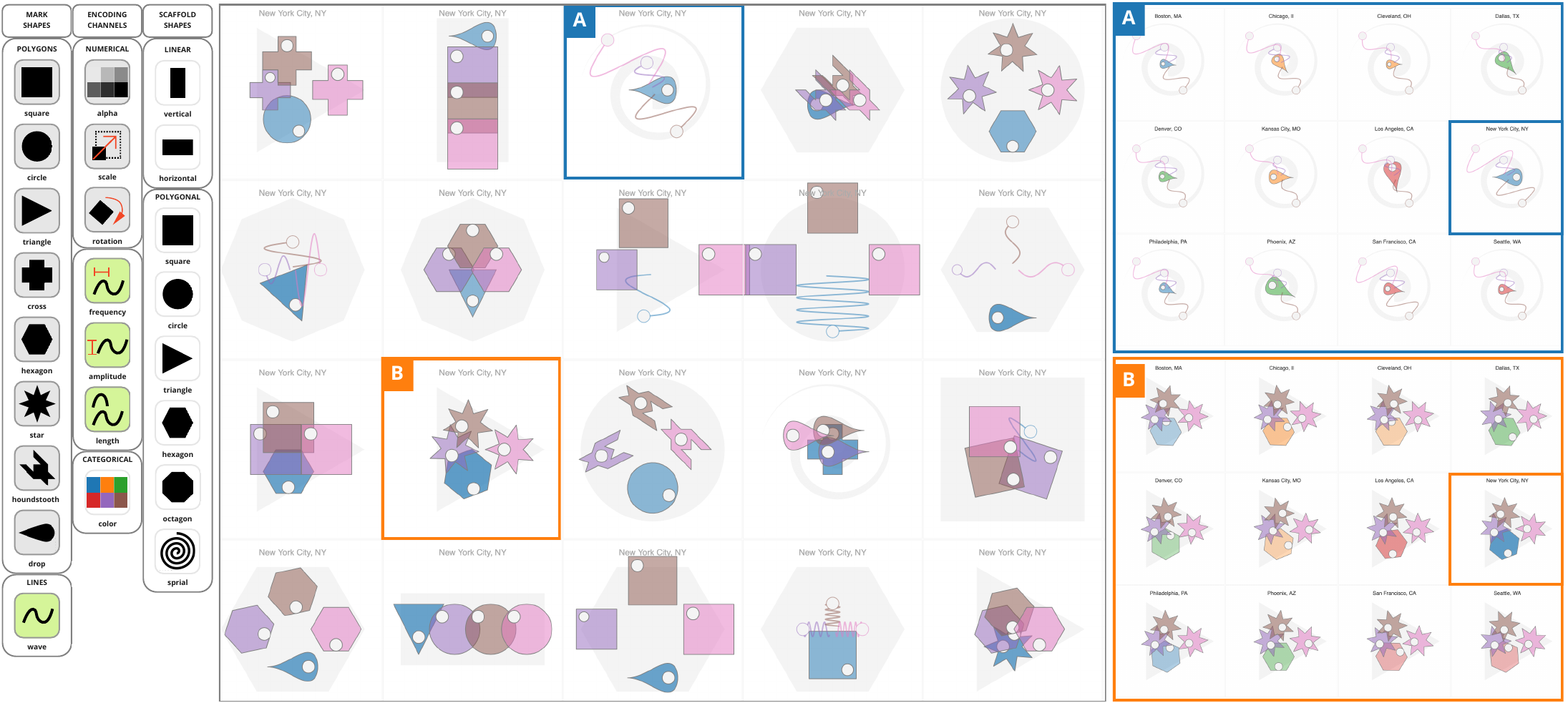}
    \caption{
    Sampling from palettes of mark shapes, encoding channels, and scaffold shapes (left), the Diatoms technique generated 20 alternative glyph designs for an urban mobility dataset~\cite{bigcities} (center), displayed as \textsl{small permutables}, in which we show a single data point (New York City) drawn according to each of the 20 designs. We highlight two of these designs as \textsl{small multiples}, where every data point is drawn the same way (right). In these two designs, a city's region, area, and population correspond respectively with the drop's fill color, size, and rotation in \caphli{A} and with the hexagon's fill color, rotation, and alpha level in \caphlii{B}; a city's bike, transit, and walk scores correspond respectively with the purple, brown, and pink waves' amplitudes in \caphli{A} and with the stars' rotations in \caphlii{B}.}
  \label{fig:teaser}
  
}

\vgtcinsertpkg

\begin{document}


\firstsection{Introduction}

\maketitle

\label{sec:intro}

Inspiration for novel visualization design can come from many sources.
In this paper, we present a novel technique for drawing inspiration from the data itself, revealed through the use of a generative design process in combination with interactive design externalization.

We concentrate on design inspiration for glyphs: small visual objects comprised of multiple marks, where the visual properties of a mark correspond with values from a single data point~\cite{borgo2013glyph,fuchs2016systematic}.
Our work is motivated by the prevalence of glyphs in communicative visualization and the lack of support in existing tools for designing and constructing glyphs.
Their design involves choices about the number of distinct marks, the relative positioning of said marks, and the encoding channels used to convey data values, choices that impact how viewers visually discriminate the glyphs.
However, glyph design is not solely about perceptual concerns. 
Designers also consider aspects such as visual symmetry, the presence of emergent patterns and figurative associations, as well as how these aspects interact with the semantics of the underlying data. 
These aspects of glyph design are difficult to specify a priori.
In light of these difficulties, we allocate glyph generation to a constrained sampling process, one capable of producing a continuous sequence of candidate glyph designs, whereupon the designer becomes a curator~\cite{gross2018generative}, tasked with identifying promising designs and excluding less promising ones based on their own abstract and ineffable criteria.  

Our primary contribution is \textit{Diatoms}, a technique that encapsulates the defining characteristics of generative design~\cite{gross2018generative} (repetition, randomness, and logic) to provide glyph design inspiration by sampling from palettes of mark shapes, encoding channels, and glyph scaffold shapes (\autoref{fig:teaser}-left).
To review and navigate between alternative glyph designs, our sampling process is supported by two modes of interactive design externalization.
The first is a familiar \emph{small multiples} configuration in which every data point is drawn according to the same design specification (\autoref{fig:teaser}-right), coupled with the ability to page between alternative glyph designs. 
The second mode is what we refer to as a \emph{small permutables} design gallery, in which a single data point is drawn according to each of the generated glyph designs (\autoref{fig:teaser}-center), coupled with the ability to page between data points.
Our secondary contributions include observations from interviews with 12 designers, to whom we demonstrated an implementation of our technique as an extension to Tableau \revision{that featured three example palettes}.
Finally, we discuss the potential of our technique \revision{in terms of} how it might be integrated into interactive authoring systems, so as to connect the process of inspiration gathering with bespoke visualization construction.

\section{Background and Prior Research}
\label{sec:rw}

We draw upon visualization research and practice as well as adjacent domains' incorporation of design externalization and generative design. 

\subsection{Inspiration for Visualization Design}
\label{sec:rw:inspiration}

Recent interactive visualization construction tools allow people to craft bespoke visualization beyond conventional statistical charts.
These tools include Lyra~\cite{satyanarayan2014lyra}, Data Illustrator~\cite{liu2018data}, Charticulator~\cite{ren2018charticulator}, and most recently StructGraphics~\cite{tsandilas2020structgraphics}.
Whether the goal is to realize \textit{mash-ups} of existing chart types or to draw \textit{xenographics} (\textit{``weird but (sometimes) useful charts''}~\cite{xenographics}), the output of these authoring tools typically serves a specific communicative intent~\cite{kosara2016presentation}, where transferability of the output to other datasets is not as critical as novelty and memorability~\cite{borkin2013makes}.
This communicative intent stands in contrast to those of other visualization construction environments~\cite{grammel2013survey} where the output is to be used for analyzing data and should generalize across datasets and use cases.
A common critique of bespoke visualization construction tools~\cite{satyanarayan2019critical} is that they are authoring tools, not design tools, in that they do not provide any design inspiration or support.
These tools assume that people already have a particular design in mind when approaching these tools; absent a preconceived design, they face a blank canvas.
With Diatoms, we address this missing step in visualization construction by providing design inspiration via a sampling-based process coupled with a comparative display of design alternatives, albeit with a focus on glyph-based visualization.
We distinguish inspiration from recommendation, in that we associate the former with a desire to produce a novel visualization to support a communicative intent while the latter addresses analytical intents, exemplified by projects like Show Me~\cite{mackinlay2007showme} and Voyager~\cite{wongsuphasawat2017voyager}.

\bstart{Inspiration from others}
Sources of visualization design inspiration vary across individuals and communities of practice.
The public-facing work of others is one such source, particularly given the volume of work appearing in news media, at practitioner conferences, within visualization communities on Twitter and Reddit, or on the \textit{\#share-inspiration} channel of the Data Visualization Society's Slack~\cite{dvs}.
Some seek inspiration among others working with the same tools or languages.
For example, D3.js developers can find inspiration on indexed repositories such as Bl.ock Builder~\cite{BlockBuilder} or bl.ocksplorer~\cite{Blocksplorer}, while among the Tableau community, people seek out inspiration on Tableau Public~\cite{tableaupublic}.
While it is possible to download or fork specific example implementations from these repositories or extract elements from published vector-based charts with tools like SVG Crowbar~\cite{crowbar}, two issues arise:
first, differences between the dataset used in an example implementation and the dataset at hand can hinder the use of the example as a starting point for design;
second, the new design can be derivative of the original on which it was based.

\bstart{Visual metaphors}
Another source for visualization design inspiration are visual languages used in other media, from abstract art to musical notation and engineering diagrams~\cite{Lupi2018OCD}.
The natural world is also a boundless source for visualization design inspiration: we point to examples of botanical motifs evoking trees~\cite{cruz2019dendrochronology,kleiberg2001botanical} and flowers~\cite{stefaner2017}, geological motifs evoking sedimentary layers~\cite{Viegas_InfoVis_2004}, celestial motifs evoking constellations~\cite{bremer2016}, or biomimetic motifs~\cite{eggermont2018bio} evoking the behavior of flocks or swarms~\cite{aseniero2016fireflies}.
Even the human face~\cite{chernoff1973use} and body~\cite{horn1998metaphor} have served as inspiration for visualization design.
However, whether drawing inspiration from other visual media or from the natural world, the structure and distributions of values in the dataset at hand may not be congruent with a particular visual metaphor.
In light of this, our approach attempts to generate visualization design inspiration independent of external influences.

\lettrine[lines=6,findent=2mm,nindent=-.5mm]{\includegraphics[width=33mm]{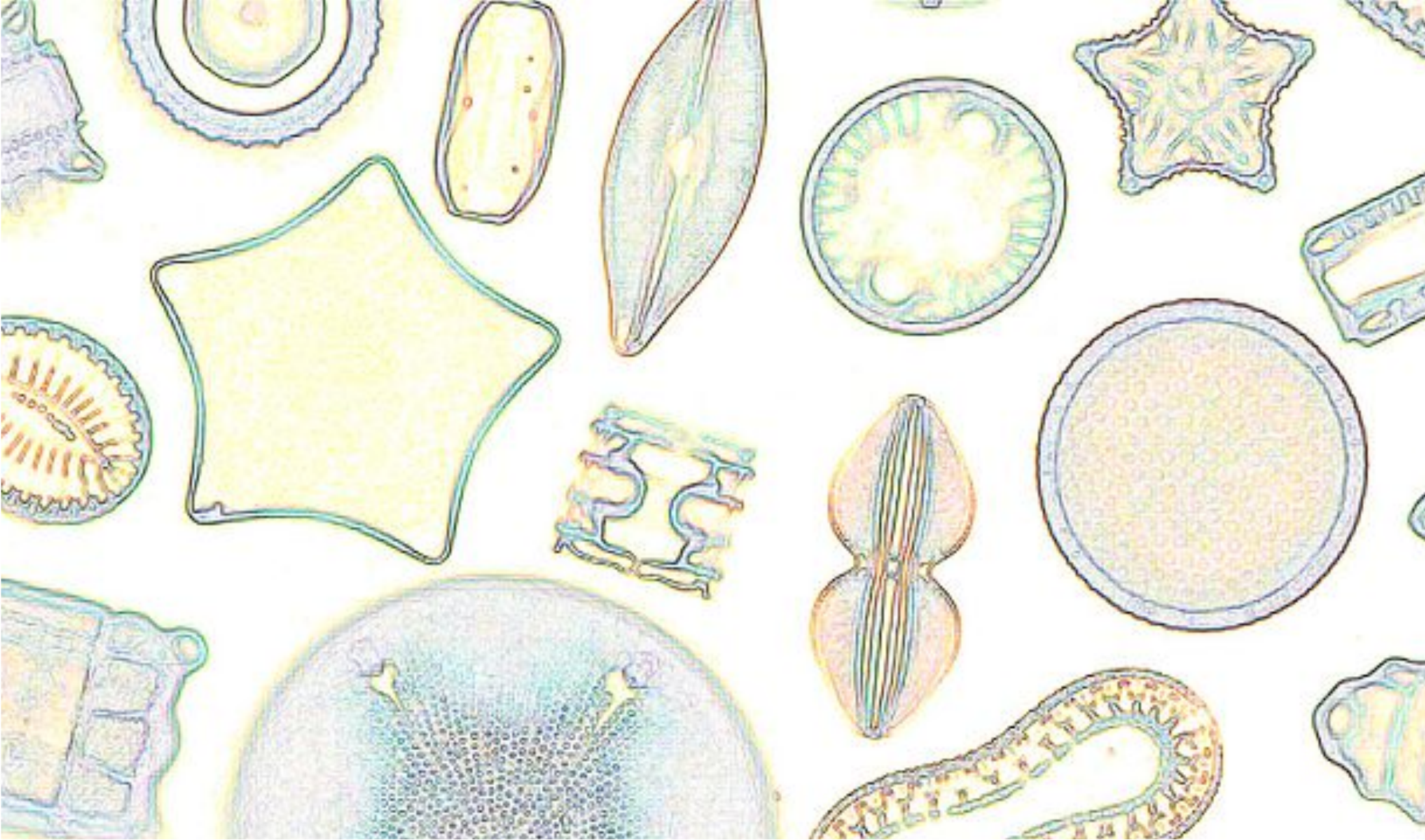}} 
As we discuss in Sections~\ref{sec:reflection} and~\ref{sec:study:figurative}, our sampling-based approach does not preclude the serendipitous recognition of visual motifs found in nature or other media.
In our case, the naming of the Diatoms technique came about when our approach generated patterns reminiscent of the eponymous microscopic algae, such as those depicted in the inset image at left~\cite{diatoms:wikimedia}.

\subsection{Glyph-Based Visualization}
\label{sec:rw:glyph}

While new techniques and sources for design inspiration are needed for all forms of visualization, we restrict our scope to glyph-based visualization.
In a 2013 survey, Borgo~\etal~\cite{borgo2013glyph} describe a glyph as a \textit{small visual object that depicts attributes of a data record}.
Fuchs~\etal~\cite{fuchs2016systematic} elaborated upon this definition, describing glyphs as instances where \textit{single data points are encoded individually by assigning their dimensions to one or more marks and their visual variables}. 

\bstart{The use of glyphs in practice}
Our scope reflects the prevalence of glyphs in bespoke communicative visualization by practitioners; we point to examples in information design award showcases~\cite{andrews2014,stefaner2014}, in visualization trade journals ~\cite{jhunja2021,ruzicka2020}, in celebrated collections like \textit{Dear Data}~\cite{Lupi2016DearData} or \textit{Data Sketches}~\cite{bremer2016b,wu2017}, or in featured collections on Tableau Public~\cite{jones2020,kovacs2018}.
We are also motivated by the versatility of glyphs.
Though often arranged in grid configurations, glyphs also appear in variants of other charts, with examples appearing in embellished bar charts~\cite{stefaner2017}, maps~\cite{wickham2012glyph}, tile maps~\cite{wsj:redblueamerica}, and node-link diagrams~\cite{accurat2013}. 
The opportunity for higher information density with glyphs also allows for their use in compact arrangements in tables~\cite{perin2014revisiting}, as inline word-scale elements in bodies of text~\cite{beck2017word,goffin2014exploring}, or in small display contexts, such as the popular iOS Activity app~\cite{Activity2020} and its glyphs comprised of three concentric ring marks.

\bstart{Drawing glyphs}
Despite their prevalence and versatility, glyphs are tedious to construct using interactive tools.
For instance, Tableau experts have described how they iterate between multiple tools~\cite{SchulteKovacs2019}, echoing findings from prior interview studies~\cite{bigelow2016iterating,parsons2020}; they draw mark shapes in applications like Illustrator~\cite{illustrator} or Procreate~\cite{procreate}, transform vector graphics with tools like co\"ordinator~\cite{coordinator}, and impute and densify data with spreadsheet or scripting tools.
The tedium and difficulty of glyph construction persists among bespoke visualization authoring tools; to create a small multiples arrangement of glyphs in Charticulator~\cite{ren2018charticulator}, one must undergo a recursive process of exporting a single glyph and importing this \textit{nested chart} in a new Charticulator instance\cite{satyanarayan2019critical}. 

Difficulties associated with constructing glyphs also reflects the immensity of their design space.
\revision{Beyond the initial selection of mark shapes and encoding channels, designers must be conscious of how the perception of mark shape, fill color, and mark size can interact~\cite{smart2019measuring}, and they must weigh the effects of juxtaposing or superimposing marks on Gestalt perception within and between glyphs.}
Accordingly, there are varying approaches to glyph design, including experimenting with visual metaphors from nature or other media~\cite{Lupi2016DearData}, appropriating semantically-related figurative elements and frames~\cite{byrne2019figurative,kim2016data}, and constructing taxonomies of abstract mark arrangements~\cite{maguire2012taxonomy}.
With our sample-based approach to glyph design, we relieve designers of these decisions, at least early in the design process, asking them to instead act as a curator of generated designs. 

\subsection{Design Externalization}
\label{sec:rw:externalization}

Spatially juxtaposing alternative designs is common across visual disciplines.
Scholars studying the history of design employ this technique to study variations across designers, historical periods, and geography, facilitating conversation among collaborators and inspiring new research questions~\cite{davis2019design}.
Meanwhile, designers collect and visually juxtapose design inspiration produced by others, whether on physical walls or virtual whiteboards.
In this paper, we are concerned with designers employing this technique to discuss and critique their own designs.

\bstart{Automating design externalization}
While it is possible to arrange design variants manually, automated design externalization can facilitate in-the-moment comparison of variants as they are produced.
This automation is well-suited for drawing and photo editing applications~\cite{lee2010designing,marks1997design,terry2002side}, where an interactive parameter space can quickly produce many variants.
Interactive design externalization can facilitate browsing and presenting design alternatives~\cite{buxton2000large} as well as a sequential exploration of a parameter space~\cite{koyama2020sequential}, either by individuals or by teams of designers engaged in a process of collaborative critique~\cite{okuya2020investigating}.

\bstart{Externalization for visualization design}
In the context of visualization, there are precedents for juxtaposing a sequence of visualization artifacts in analysis tools for the purpose of analytical provenance and auditing~\cite{heer2008graphical}, but examples of externalization in visualization design are less common. 
\revision{One precedent is the} process undertaken by the news graphics team of the \textit{The New York Times}~\cite{corum2014}, where an automated script generates a screenshot with every git commit to a project, appending this screenshot to a thumbnail gallery of visualization design variants.
While this approach may not easily afford quick comparisons of designs with minute differences or of charts that are ideally suited for full-screen viewing, it may be appropriate for comparing alternative designs of small glyphs.
\revision{Another precedent for externalization in visualization design can be found in Schroeder and Keefe's Visualization-by-Sketching interface~\cite{schroeder2015visualization}, wherein a designer can vet a candidate mark design by interactively generating a gallery that illustrates how the mark may manifest in different conditions, such as by varying mark density and size.
We incorporate ideas similar to both precedents in Diatoms, from comparing across glyph design choices to comparing alternative sizes and positions of a particular glyph design. 
}

\subsection{Generative Design}
\label{sec:rw:generative}

A common characteristic of conventional design is the step-by-step and often manual process of realizing a new idea.
In visualization design, this captures the trial-and-error process of associating visual encoding channels to attributes of data, which could be informed by an understanding of graphical perception or by domain conventions.
In contrast, generative design suggests an alternative approach that fixates on the identification of configurable parameters and the refinement of rules for generating \textit{many} ideas and especially those that exhibit emergent yet desirable traits that cannot be easily specified beforehand. 
With respect to glyph-based visualization, these traits could include strong Gestalt associations or memorable figurative associations.  
In generative design, Gro{\ss}~\etal~\cite{gross2018generative} describes how \textit{traditional craftsmanship recedes into the background, and abstraction and information become the new principal elements}.
The formalization of abstract rules typically involves elements of repetition, logic, and randomness. 
Artists and designers have applied generative processes in many domains; we point to examples in interface design~\cite{swearngin2020scout}, architecture~\cite{goldsteinspaceanalysis}, urban planning~\cite{delve}, industrial design~\cite{matejka2018dream}, and even music~\cite{eno1996}.
Despite its prevalence in these fields, generative design is under-utilized in visualization research.

\bstart{Generative design for visualization}
Data art and data visualization are ideal application scenarios for generative design~\cite{datastories}.
The visual nature of the artifacts produced allow for repetition to manifest both temporally and spatially, the latter being conducive to externalization of alternative designs in design galleries~\cite{lee2010designing,marks1997design}.
Designers and artists must consider how to balance randomness with logical decisions that bind data types and values with visual properties.

One noteworthy recent example of interactive design externalization coupled with a generative process is Morph~\cite{morph2018}, a tool for generating visual art from tabular data; Morph seeds the design space with a familiar statistical chart and applies random \textit{mutations} to visual encoding channels, resulting in a visual branching of increasingly mutated designs, which fans outward with additional selections.
Like Morph, we also embrace randomness to an extent with Diatoms, sampling from encoding channel palettes for each mark in a glyph.

\begin{figure*}[t!]
  \centering
  \includegraphics[width=\linewidth]{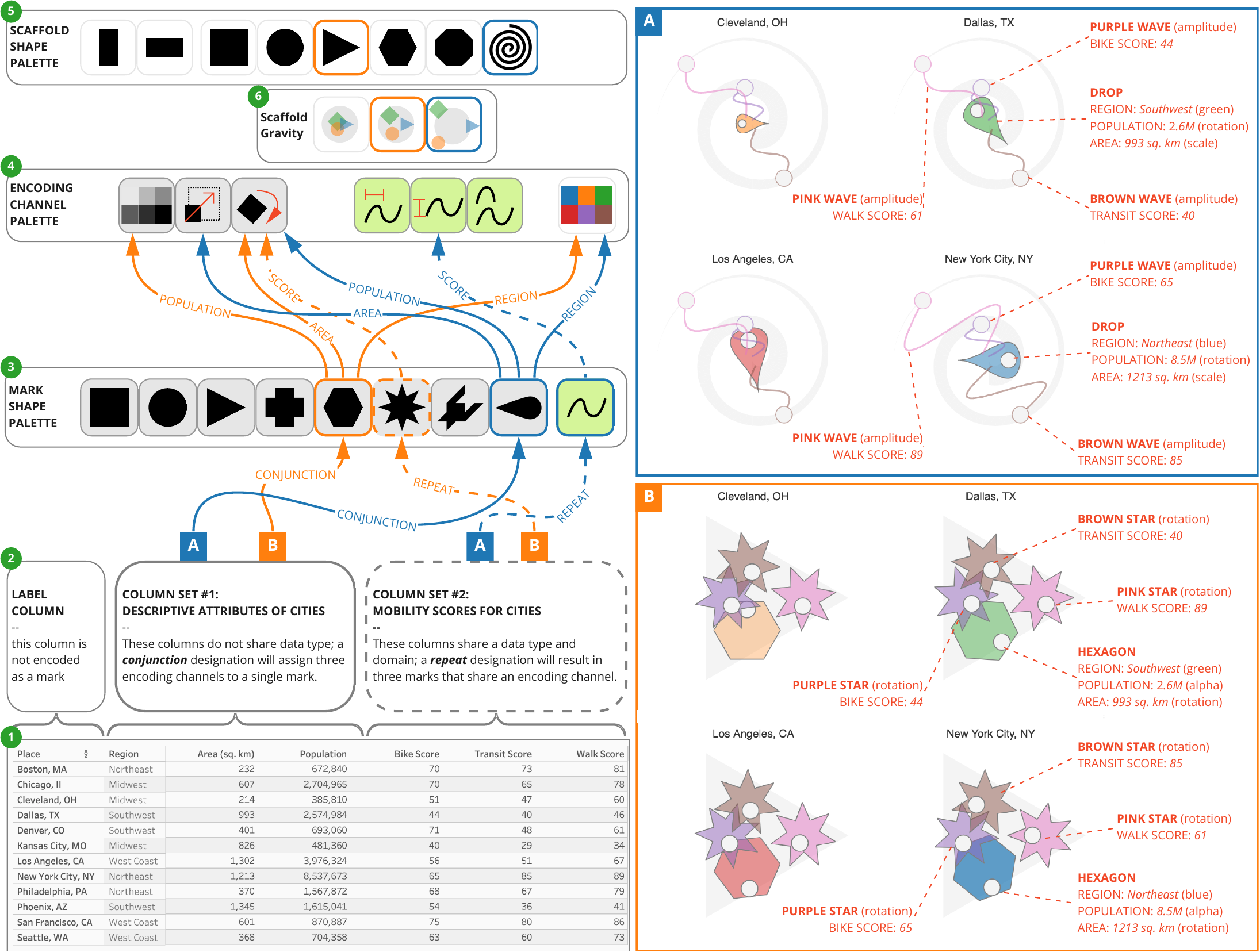}
  \vspace{-5mm}
  \caption{This diagram depicts a column set \revision{designation} (2) for a data table~\cite{bigcities} (1), as well as sampling outcomes with respect to mark shape (3), encoding channel (4), scaffold shape (5), and scaffold gravity (6) for alternative glyph designs \caphli{A} and \caphlii{B}. On the right, we added the red annotations to cropped sections of the corresponding \textsl{small multiples} configurations from \autoref{fig:teaser} (right) to highlight the encodings and underlying values.}
  \vspace{-5mm}
  \label{fig:explainer}
\end{figure*}

\section{The Diatoms Technique}
\label{sec:diatoms}

The purpose of Diatoms is to quickly generate design inspiration for glyphs, and this generation process is coupled with interactive design externalization for comparing promising designs.
We break down the Diatoms technique into a process of designating \textit{conjunction} and \textit{repeat} associations in the data, sampling from shape and channel palettes to generate alternative glyph designs, and finally comparing and curating these designs in \textit{small multiples} and \textit{small permutables} viewing modes.
Throughout this section, we refer to \autoref{fig:explainer} and an example featuring an urban mobility dataset~\cite{bigcities} to explain aspects of the technique, while the supplemental video~\cite{supplemental} illustrates its interactive aspects.

\label{sec:diatoms:semantic}

We assume tabular data expressed in wide format such as in \autoref{fig:explainer}.1.
Given such a table, Diatoms will draw a glyph for each row.
We first decide which columns to include in the glyph designs, and we optionally group columns that share semantic associations into \textit{column sets}.
Otherwise, each set contains a single column.

\subsection{Sampling from Shape and Encoding Palettes}
\label{sec:diatoms:sampling}

\begin{figure}[h!]
  \centering
  \includegraphics[width=\linewidth]{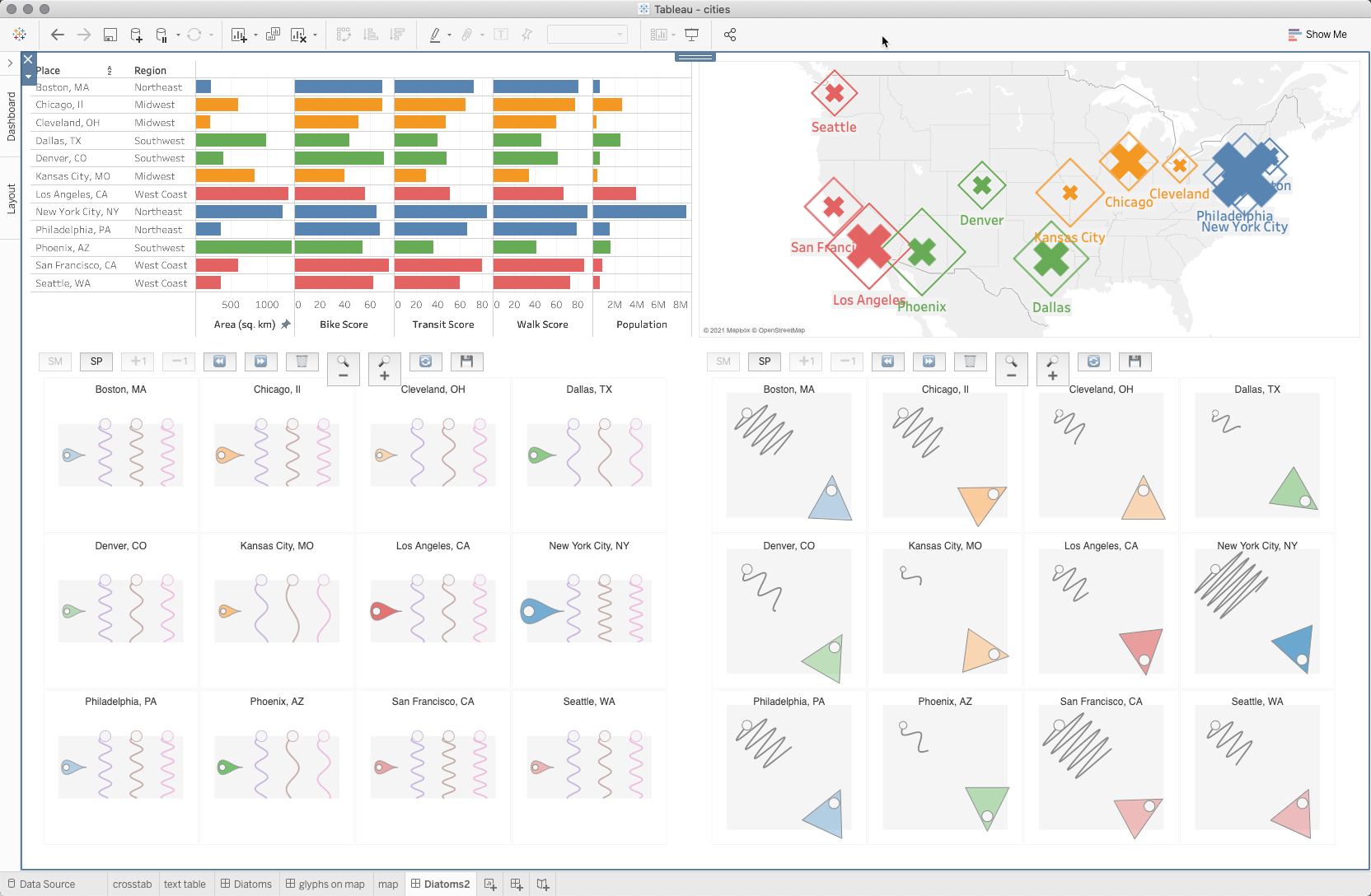}
  \vspace{-5mm}
  \caption{A Tableau dashboard containing two worksheets (top) and two instances of Diatoms (bottom), both in \textsl{small multiples} mode. A repeat designation for mobility scores is used in the bottom left instance.}    
  \vspace{-5mm}
  \label{fig:dashboard}
\end{figure}

We assign mark shapes and encoding channels via a constrained random sampling process.
The \revision{example} shape and channel palettes \revision{featured in Figures \ref{fig:teaser} \-- \ref{fig:dashboard}} represent one possible starting point in the glyph design space, others are certainly possible and \revision{should be left to the discretion of the designer}.
In \autoref{sec:reflection}, \revision{we reflect on our experimentation with palette options that preceded those featured in this section.} 

\bstart{Mark shape palette}
We assign a unique shape to each column set, and each shape can only be assigned once within a single glyph design.
Our \revision{example} shape palette shown in \autoref{fig:explainer}.3 contains eight \textit{polygon} shapes and one \textit{wave} shape, meaning that a single glyph design can contain at most nine unique mark shapes. 
\revision{Our inclusion of the \textit{wave} serves as a counterpoint to the more salient filled polygons. 
We chose a sine wave following experimentation with straight line marks, where we determined there to be an insufficient number of salient encoding channels compatible with the latter, whereas the former provided us with \textit{frequency} and \textit{amplitude} in addition to \textit{length}.
Other wave shapes are worth considering for future palettes, such as square and sawtooth waves. Most of the} polygon shapes in this palette are symmetric \revision{and familiar, reminiscent of the mark shape palettes in Tableau}.
\revision{Since we include mark rotation in our example encoding channel palette, we introduced asymmetry to this mark shape palette in two ways: by including \textit{houndstooth} (\large{\mHoundstooth}\normalsize) and \textit{drop} (\large{\mDrop}\normalsize) shapes and by adding a directional \textit{pip} to all marks: a small white circle to indicate the rotation of a shape, akin to a level indicator on a physical dial.}

\bstart{Encoding channel palette}
We also assign encoding channels to columns via sampling.
Given the small number of options in this \revision{example} palette (shown in \autoref{fig:explainer}.4), we allow for channels to be sampled by more than one column set.
We include one categorical channel (\textit{mark color}) and \revision{an equal number of} quantitative channels for polygons (\textit{alpha, size, rotation}) and waves (\textit{frequency, amplitude, length}).
\revision{In \autoref{sec:reflection}, we discuss our prior experimentation with several other encoding channels (\eg position, distortion, stroke properties), along with rationale for why we omitted them from this palette.}

\bstart{Scaffold shape palette}
\revision{Finally, each glyph design is randomly assigned a scaffold shape.
Our example palette (\autoref{fig:explainer}.5) has two \textit{linear} and six \textit{polygon} shapes}; 
\revision{the latter includes a \textit{spiral} as a contrast to the simple and symmetric shapes that fill out the rest of the palette.}
Scaffolds are organizing principles for marks, which are placed at equally-spaced intervals along a scaffold following column set order. 
We additionally randomize the distance of marks from the periphery of the scaffold, which we refer to as a scaffold's \textit{gravity} (\autoref{fig:explainer}.6).
Design \hli{A} has a \textit{spiral} scaffold with \textit{weak gravity}, while design \hlii{B} has a \textit{triangle} scaffold with \textit{medium gravity}.

\begin{figure*}[t!]
  \centering
  \includegraphics[width=\linewidth]{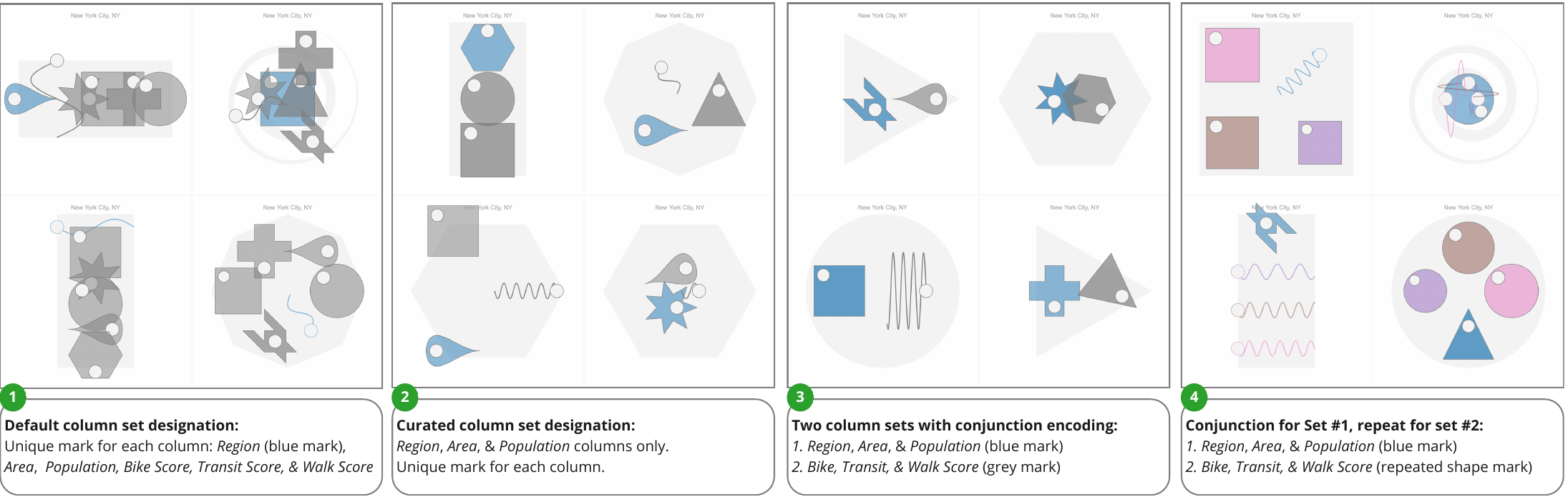}
  \vspace{-5mm}
  \caption{Variations on a column set designation with the urban mobility dataset~\cite{bigcities}, with four glyph designs shown for each designation in a \textsl{small permutables} configuration Variant 4 uses the column set designation used in \autoref{fig:explainer}.}
  \vspace{-5mm}
  \label{fig:columnsets}
\end{figure*}

\bstart{Column sets}
The cardinality and type of a column set determines the number and appearance of marks that get drawn.
\autoref{fig:columnsets} shows variations on a column set designation, with four alternative glyph designs produced for each.
Designating one column per set may not yield interpretable designs after more than a few columns, such as in the example of \autoref{fig:columnsets}.1.
While one solution is to use only a subset of columns, such as in \autoref{fig:columnsets}.2, we can also combine columns into \textit{sets}.
The example designation in \autoref{fig:explainer}.2 features two columns sets, and these sets represent two types of associations between columns. 

\istart{Column sets with \textbf{conjunction} designations}
The first column set contains descriptive aspects of cities: \textsc{region}, \textsc{area}, and \textsc{population}.
This set will correspond with a single mark exhibiting a \textit{conjunction} encoding~\cite{ware2004information}, where each column corresponds with a unique encoding channel.
In glyph design \hli{A}, this set is assigned a \textit{drop} shape, while it is assigned a \textit{hexagon} shape in glyph design \hlii{B}.
For the \textit{drop} mark in glyph design \hli{A}, \textsc{region}, \textsc{area}, and \textsc{population} are respectively assigned to \textit{color}, \textit{size}, and \textit{rotation}.
For the \textit{hexagon} mark in glyph design \hlii{B}, these columns are assigned to \textit{color}, \textit{rotation}, and \textit{alpha level}.
The blue marks in Figures~\ref{fig:columnsets}.3 and~\ref{fig:columnsets}.4 are alternative outcomes of this designation.

\istart{Column sets with \textbf{repeat} designations}
\textsc{bike score}, \textsc{transit score}, and \textsc{walk score} share a common numerical domain from 0 to 100; combining them in a column set with a \textit{repeat} designation means that each column will be assigned a unique mark, but each mark will \textit{repeat} a shared shape and channel combination, so as to promote a direct comparison of values following the Gestalt principle of similarity.
For a column set with a \textit{repeat} designation, we distinguish the set's columns using color.
Unlike the colors corresponding with \textsc{region} in the first column set, these colors do not reflect categorical values appearing in table cells.
This \textit{repeat} designation could be thought of as a pivot transformation on these columns, from a wide data format to a long one, where column names become categorical values amenable to encoding with a channel such as color.
Color assignments are unique, meaning that the combined sum of categories across categorical columns and the number of columns appearing in sets with \textit{repeat} designations should be fewer than the number of distinguishable colors. 
In glyph design \hli{A}, the \textit{amplitude} of the purple, brown, and pink \textit{wave} marks correspond with the three mobility scores, while in glyph design \hlii{B}, they correspond with the \textit{rotation} of the \textit{star} marks.
The grey marks in \autoref{fig:columnsets}.3 and the repeated marks in \autoref{fig:columnsets}.4 illustrate the difference between a \textit{conjunction} and \textit{repeat} designation for these columns, with the latter being the same designation used in \autoref{fig:explainer}.

\subsection{Small Multiples and Small Permutables}
\label{sec:diatoms:externalization}

Given a column set designation, sampling across the three palettes can generate many possible glyph designs, and the number of possible permutations of mark shape, channel, and scaffold shape grow as designations include more columns.
We therefore propose two ways to view and navigate between alternative designs. 
The first viewing mode is a familiar \textit{small multiples} grid configuration in which a glyph for each data point is drawn according to the same glyph design, such as in the two grids of 12 glyphs for designs \hli{A} and \hlii{B} in \autoref{fig:teaser} (right).
In our implementation of Diatoms, we provide the means to page between small multiples configurations of alternative glyph designs.
The second mode is a design gallery of glyph alternatives which we refer to as a \textit{small permutables} grid configuration, in which one data point is drawn multiple times, once for each alternative glyph design, such as in \autoref{fig:teaser} (center).

Whereas navigation in \textit{small multiples} mode pages between alternative designs, in \textit{small permutables} mode it pages between data points. 
Clicking on a glyph selects it, which serves to preserve context when toggling between the two viewing modes.

\bstart{Curation}
Diatoms seeds the inspiration process with an initial set of alternative glyph designs.
In our current implementation, we generate five as a starting point. 
At any time, we can generate and append additional glyph designs as desired, or we can cull uninspiring ones.

\bstart{\revision{Legends on demand}}
In the supplemental video~\cite{supplemental}, we show how mousing over a glyph reveals a tooltip-based legend indicating the correspondences between marks and data values, a consolidated version of the red annotations added to \autoref{fig:explainer} (right).
\revision{Legends are critical for glyph-based visualization, particularly when glyphs are presented to their intended viewing audience. 
Legends may be less critical for rapid design iteration, where unpromising glyph designs can be dismissed without scrutinizing their legends.
Our current approach is a compromise.
On the one hand, a designated and always-visible legend would be compatible with our \textit{small multiples} viewing mode, where every glyph shares a design. 
On the other hand, our \textit{small permutables} viewing mode would require a unique legend for every glyph design; an always-visible legend for every design, shown adjacent to each design or shown as annotations akin to \autoref{fig:explainer} (right) would result in excessive visual clutter, impeding design comparison and iteration.}

\begin{figure}[h!]
  \centering
  \includegraphics[width=\linewidth]{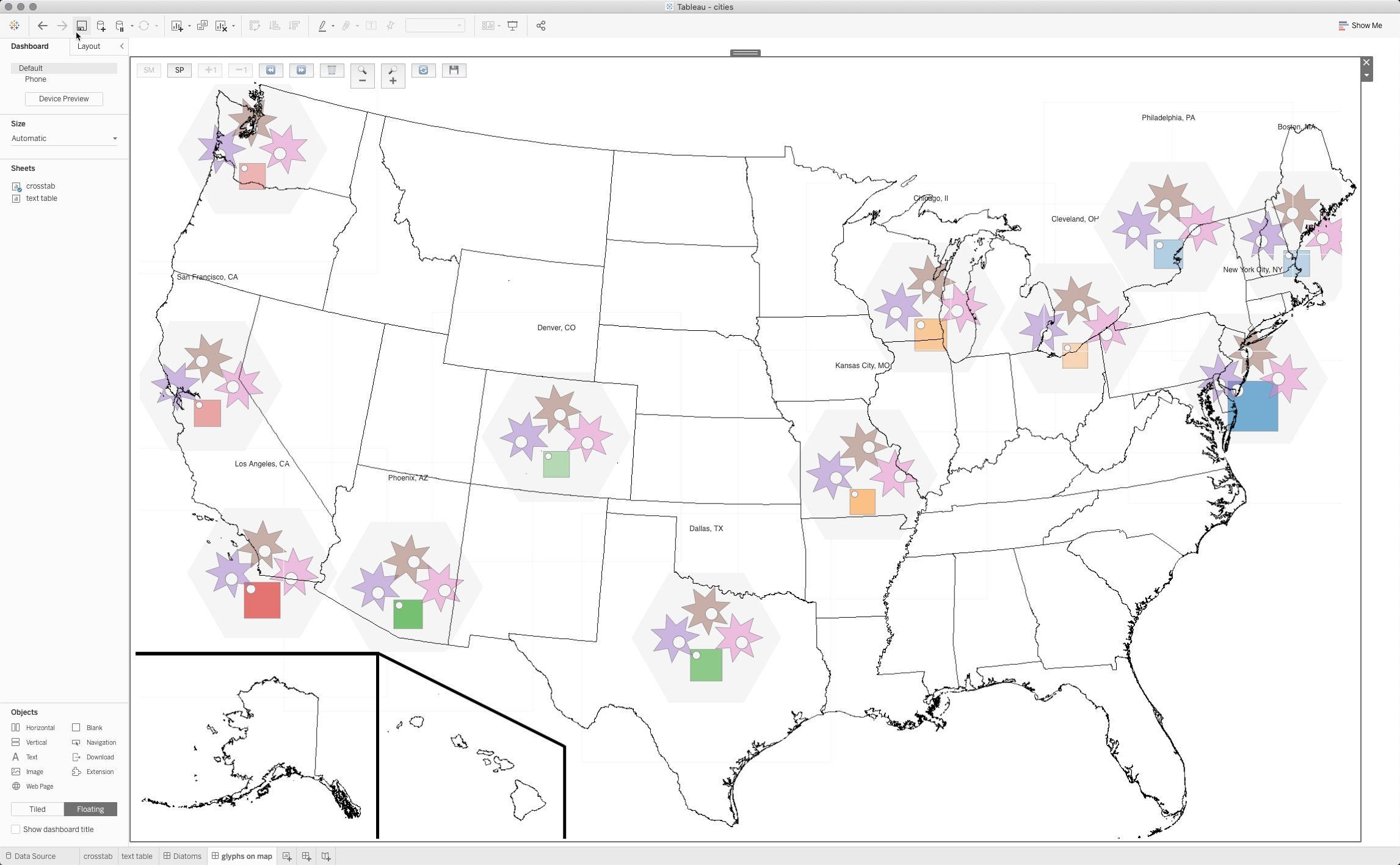}
  \vspace{-5mm}
  \caption{Resizing and arranging glyphs allows designers to assess the viability of a particular design for manifestations other than a \textsl{small multiples} grid configuration. In this example, a set of glyphs are manually arranged in reference to a map image in Tableau Desktop.}    
  \vspace{-1mm}
  \label{fig:glyphmap}
\end{figure}

\bstart{Resizing and positioning \revision{candidate} designs}
Beyond navigation and curation, we anticipate other ways in which designers may want to assess alternative glyph designs.
First, viewing the glyphs at different sizes allows us to determine a particular design's suitability for different viewing contexts, such as placing glyphs within a table or a small display context.
Second, we may want to break the grid to view the glyphs according to a customized spatial configuration; any selected glyph can be dragged to a new location in the canvas, and when used in conjunction with glyph resizing, this functionality could be used to assess the viability of glyphs for placement in a scatterplot, a tile map, or a symbol map such as in \autoref{fig:glyphmap}.
In \textit{small permutables} mode, this spatial reconfiguration could be used to group or rank alternative designs, so as to compare a subset of promising designs side by side.

\subsection{Implementation of Diatoms}
\label{sec:diatoms:realization}

We demonstrate an implementation of Diatoms in Tableau Desktop~\cite{tableau} via its Extensions API~\cite{tableauext}.
This choice of implementation allowed us to defer the tasks of data shaping, data type inferencing, and exploratory analysis to Tableau, such that Diatoms only ingests a tidy data table that requires no further filtering or aggregating.
Finally, this implementation context allows for an instance of Diatoms to be combined with other Tableau content in a dashboard, including other instances of Diatoms, as depicted in \autoref{fig:dashboard}. 
We implemented Diatoms in JavaScript and used p5.js~\cite{p5js} to render the glyphs to a canvas element.

\begin{figure*}[t!]
  \centering
  \includegraphics[width=\linewidth]{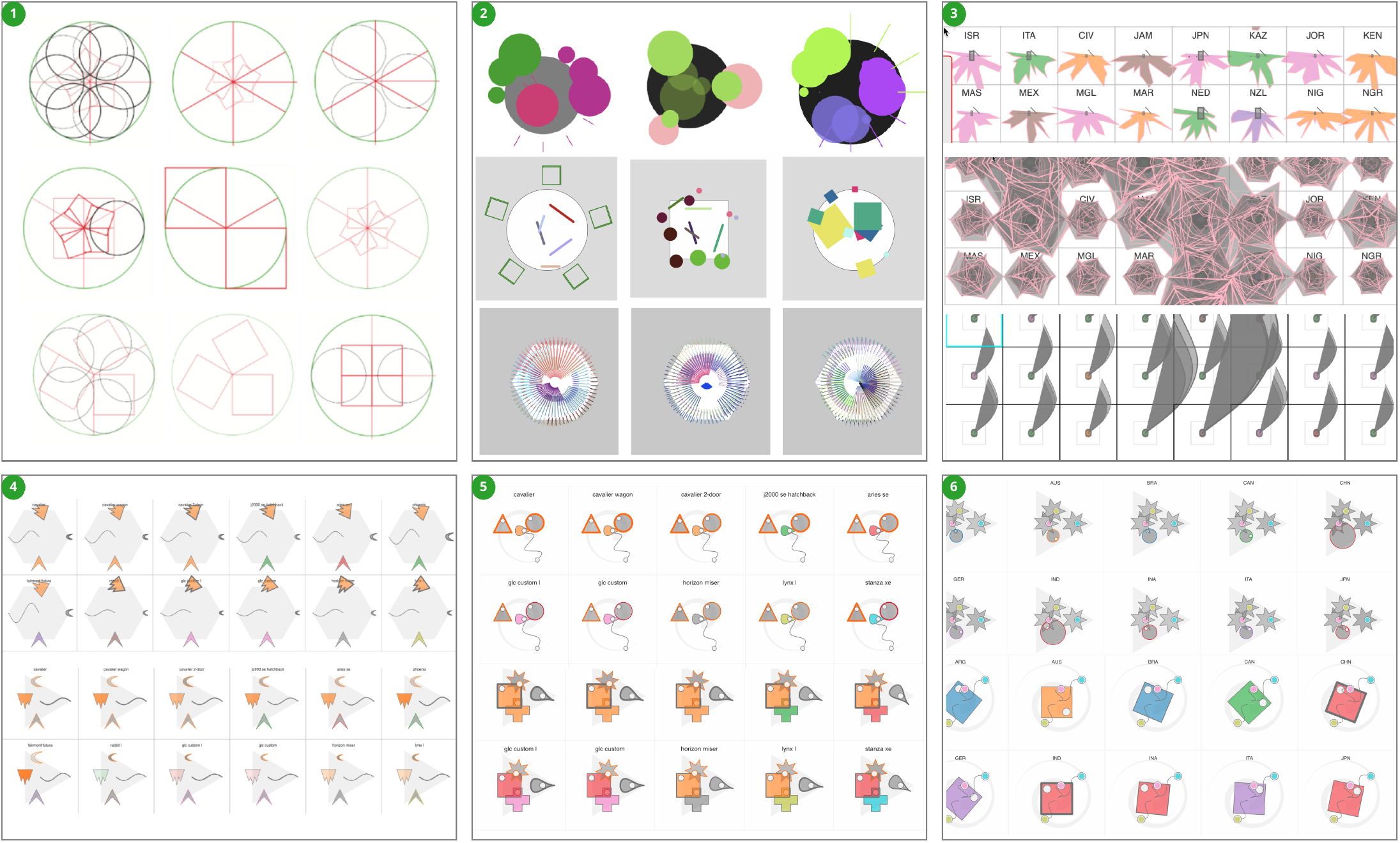}
  \vspace{-5mm}
  \caption{The evolution of the Diatoms technique, including experiments with: scaffold and mark permutations (1); position- and repetition-based encodings with a small palette of mark shapes (2); parameterized shape encodings (3); mark shape palettes comprised predominantly of asymmetric shapes (4); mark stroke color and weight encodings (5); and colored \textsl{pips} for marks corresponding with column sets having repeat designations (6).}
  \vspace{-4mm}
  \label{fig:progress}
\end{figure*}

\section{\revision{Technique Evolution and Palette Experimentation}}
\label{sec:reflection}

We reflect on our experimentation with generative processes and on the alternative \revision{mark shape, encoding channel, and scaffold shape palettes that preceded the example palettes featured in Figures \ref{fig:teaser}--\ref{fig:dashboard}}.
\autoref{fig:progress} contains snapshots from our experimentation, encompassing a period of six months, during which time the authors met weekly to critique the generated designs and the palettes that we sampled from.

\bstart{Scaffold and mark permutations}
Inspired by the prevalence of glyphs with marks arranged in circular or radiating patterns (\autoref{sec:rw:glyph}), we initially experimented with generating permutations of circular scaffolds \revision{populated using a simple palette of circle, square, and line marks}. 
Even before we bound data to these marks, arrangements of placeholder marks such as those shown in \autoref{fig:progress}.1 conveyed the variety of possible designs, even with a small number of marks. 

\bstart{Position and repetition as encoding channels}
The glyph designs shown in \autoref{fig:progress}.2 capture our experimentation with central and peripheral mark positioning and non-circular scaffolds.
We also assessed position-based encodings, where data values would determine the position of marks relative to the scaffold.
However, anticipating the potential deployment of glyphs in contexts where the position of the glyph itself is meaningful, such as on map like in \autoref{fig:glyphmap}, we \revision{omitted intra-glyph position from subsequent encoding channel palettes} in favor of an even spacing of marks relative to their scaffold.

\bstart{\textit{Small multiples} and parameterized shape encodings}
Though displayed in a grid formation, we generated the scaffolds in \autoref{fig:progress}.1 and the glyphs in \autoref{fig:progress}.2 from small p5.js \textit{sketch} programs~\cite{p5editor} that afforded a serial comparison of designs.
Furthermore, these sketch programs used a single example data point to generate one design per run cycle.
Realizing the need to compare glyph designs across data points, we began testing ideas in \textit{small multiples} configurations using Tableau as a backend, which allowed us to test multiple data points and a variety of datasets. 

\autoref{fig:progress}.3 shows our experimentation with parameterized shape encodings, such as by modulating the position and number of vertices in a polygon or the parameters of curves.
This led to the emergence of shapes exhibiting degrees of \textit{`spikiness'} and those evoking patterns of malignant growth, where glyphs would infringe upon the space of their neighbors.
We found shape modulation encodings to be visually dominant, diminishing the salience of other marks and encoding channels present in a glyph.
Moreover, the underlying data values driving the modulation of shapes were not readily apparent, leading us to \revision{omit shape modulation from subsequent encoding channel palettes}.

\bstart{Asymmetric marks and figurative associations}
The use of rotation as an encoding channel prompted us to consider palettes comprised predominantly of asymmetric shapes.
Though these palettes yielded memorable designs and figurative associations, such as birds and fish heads in \autoref{fig:progress}.4, shapes such as moons and arrowheads were imbued with cultural connotations, evoking petroglyphs and other motifs unrelated to the data.
Seeing these patterns prompted us to reconsider \revision{palettes of} abstract symmetric shapes and another way to signify rotation: adding circular \textit{`pips'} to marks.
\revision{Our subsequent mark shape palettes that combined symmetric and asymmetric shapes still elicited figurative associations, albeit natural rather than cultural ones, with the former inspiring us to adopt the Diatoms moniker for our technique.}

\bstart{Mark stroke encodings}
In an effort to \revision{consider broader encoding channel palettes}, we considered mark stroke color and weight as additional categorical and numerical channels, respectively.
Though the former provided another categorical channel beyond mark fill color, \autoref{fig:progress}.5 shows how mark stroke channels were difficult to interpret; they were dependent upon the shape and size of mark, and when resizing the glyphs or viewing them in a smaller viewport, it became difficult to differentiate mark stroke widths. 

\bstart{Distinguished \textit{pips} for repeated marks}
Another \revision{way} to distinguish marks corresponding to a column set with a \textit{repeat} designation is the coloring of mark \textit{pips}, as shown in \autoref{fig:progress}.6.
As with mark stroke encodings, colored pips were insufficiently salient, leading us to color the marks themselves, as described in \autoref{sec:diatoms:sampling}.
Since color is the only categorical \revision{encoding channel in the palette featured in Figures \ref{fig:teaser}--\ref{fig:dashboard}}, this design choice means that repeated marks can only encode numerical values. 
To encode \revision{multiple} categorical values, a palette would require additional categorical channels, such as textured fill patterns or stroke styles.

\section{Interview Study}
\label{sec:study}
To better understand how Diatoms might integrate into design workflows, we conducted open-ended interviews and chauffeured demos~\cite{lloyd2011human} with twelve designers.
These hour-long interviews focused on the potential utility of generative design processes and what additional functionality designers would require for connecting design inspiration with visualization authoring. 
\revision{Our approach represents an effort to collect candid reactions from designers in the absence of an established evaluation methodology; between previous experiments assessing the perceptual efficacy of particular glyph designs~\cite{fuchs2016systematic} and design reproduction studies assessing the potential of interactive visualization authoring tools~\cite{ren2018reflecting}, we argue that existing methods cannot be used to study the early and divergent stages of visualization design.}

\bstart{Participants}
We recruited seven design students (P1--P7: 6~\faVenus, 1~\faMars) and five experienced designers (P8--P12: 1~\faVenus, 4~\faMars) who work with visualization.
The former group were enrolled in a graduate course in information design and visual cognition, and we interviewed them as they were completing an assignment in glyph design.

  

\bstart{Format}
We introduced the Diatoms concept and implementation to all interviewees prior to speaking with them. 
For the students, this took place during a guest lecture, while we sent an extended version of our supplemental video\cite{supplemental} to the professional designers. 
We began the interviews by discussing the interviewees' experience with design and visualization tools, their sources of inspiration, and their current workflows. 
We then provided a choice of three small datasets (city mobility data~\cite{bigcities}, film metadata from IMDb~\cite{imdb}, or socioeconomic data from Gapminder~\cite{gapminder}). 
\revision{After giving the interviewees an opportunity to ask us clarifying questions about their chosen dataset or the concept underlying Diatoms,} we began a chauffeured demo to elicit their observations and questions about Diatoms' output. 
\revision{For each demo, we used the same palettes as those featured throughout Figures \ref{fig:teaser}--\ref{fig:dashboard}.
We initially constrained the range of possible outputs by demonstrating the technique with fewer data columns and no set designations (such as in \autoref{fig:columnsets}.2).
After discussing this initial output, we added more data columns and introduced \textit{repeat} and  \textit{conjunction} set designations (such as in \autoref{fig:columnsets}.3--\ref{fig:columnsets}.4).}
Our interviews were seeded with a set of questions that probed into why interviewees found some glyph designs to be promising and others not, as well as how they would further refine the more promising glyph designs. 
As all interviews took place via video chat and screen-sharing, the first author `drove' Diatoms in response to interviewees' observations and critiques of specific glyph designs or their requests to generate new ones; the other authors observed and took notes.
\revision{We recorded screen capture video and audio from these interviews; the following discussion reflects our thematic analysis of the transcripts and notes from these interviews.}

\subsection{General Impressions}
\label{sec:study:impressions}

\revision{Overall,} designers and students alike were \revision{generally} receptive to our technique. 
P10's comments capture this sentiment: \textit{``conceptually,} [Diatoms] \textit{is an awesome idea, it speaks to more of the playful elements that people like experimenting with.''}

Diatoms could \textit{``speed up the process of generating ideas''} [P1], allowing designers to \textit{``do some rapid sketching,} [\ldots with] \textit{different sizes, shapes and colors right off the bat''} [P2].
P9 explained that the time typically required to experiment with encodings meant that he would not be able to assess multiple options; he saw Diatoms generate alternatives in seconds that would otherwise take hours. 

Beyond accelerating the early design phase, P3 suggested that Diatoms could also help novice designers: \textit{``this is good for someone who isn't as creative,} [\ldots] \textit{they can generate something easily and not put much thought into it''}. 
As one who does not visualize data programmatically, P3 also saw Diatoms as a way of providing designers with creative alternatives that would have previously required coding.


Finally, \revision{two interviewees described the unlikeliness of discovering a particular glyph design independent of Diatoms}.
Some mark, encoding channel, and mark placement combinations would have been otherwise overlooked because the combination suggested a violation of visualization design convention.   
For instance, a glyph with strong \textit{gravity} will bunch the marks together in the center of a scaffold and can result in partial mark occlusion; despite this, P6 noted how one such glyph design generated using the urban mobility dataset appropriately evoked the density of cities.
Even if the first impressions of a design suggest a violation of convention or a lack of perceptual clarity, P11 saw the advantage in giving these imprecise designs an opportunity to inspire.
Many designs are 
\textit{``often not clear or straightforward ideas,} [and] \textit{even though 99 are wrong; the messiness can be a virtue;''} he later added that \textit{``there are a lot of bad hits, but once in a while you stumble upon something that really works.''}




\subsection{Comparing and Winnowing Glyph Designs}
\label{sec:study:viewmodes}

In each interview, we asked about the utility of our two modes for comparing glyph designs.
The concept of \textit{small permutables} resonated with designer P8 in particular, who saw a parallel in his side practice of brand logo design, 
where he would arrange alternative logos to see which ones register and which ones should be discarded.
Upon demonstrating the ability to page through data points, he stated that this viewing mode \textit{``makes it clear how the} [values] \textit{are changing''}.
He saw \textit{small permutables} as a starting point for narrowing down one's scope, before toggling to a \textit{small multiples} mode to see all of the data points drawn according to the selected glyph design.

\bstart{First impressions can be deceiving}
\revision{We found that reactions to particular glyph designs depended on the viewing mode in which they were first encountered.} 
As only one instance of each glyph design is shown in \textit{small permutables} mode at any one time, a single design may show promise until it is applied to every row of data in \textit{small multiples} mode.
This happened multiple times to P7, who changed her mind about a design immediately upon switching modes, citing an inability to discriminate between glyphs corresponding to different data points. 
We also observed the converse reaction, where P4 did not react positively to glyph designs first seen in \textit{small permutables} mode until he saw all of the data points represented in a \textit{small multiples} configuration. 


\bstart{The best of both modes}
\revision{Upon demonstrating} both viewing modes and the ability to toggle between them, we received suggestions on how we could integrate the two modes within a single display.
While one solution could be to spatially juxtapose the two viewing modes (such as in \autoref{fig:teaser}), P11 suggested a hybrid viewing mode in which a small subset of data points are shown for each design; he specifically pointed to a section of our explanatory video in which we arranged four \textit{small multiples} screenshots within a single display, each featuring the same 12 data points drawn according to a different glyph design.
Similarly, P8 suggested a way to select a shortlist of three or four glyph designs from the existing \textit{small permutables} mode, and this selection would allow designers to \textit{`dial in'} to a more focused comparison mode, such as the one described by P11.

\subsection{Observations on Mark Shape and Channel Sampling}
\label{sec:study:palettes}

Every interviewee commented on \revision{our example palettes} of mark shapes and encoding channels.
\revision{When discussing the} various combinations assigned via sampling, \revision{some} conversations turned to steering or overriding the results of this sampling process.
For instance, P10 expressed a desire to \textit{``play around with the palettes,''} meaning a way to modulate the amount of variance in subsequently generated glyph designs, such as by weighting the sampling in favor of certain mark shapes and encoding channels. 

\bstart{Balancing symmetry, familiarity, and salience}
A recurring topic of discussion pertained to the asymmetric shapes in \revision{our example mark shape palette}.
For instance, P7 indicated that the \textit{drop} (\large{\mDrop}\normalsize) shape evoked location pins in mapping applications, particularly when rotated such that its tapered end points downwards, and this association could be misleading depending on the underlying data.
Meanwhile, P1 rightfully pointed out that the \textit{houndstooth} shape (\large{\mHoundstooth}\normalsize) \textit{``is not a common shape that people are familiar with,''} and due to its unique geometry, P2 noted that it \textit{``seems to have more meaning''} than other shapes.

The \textit{wave} shape also drew comments. 
As the sole non-polygon shape in \revision{our example mark shape palette}, both P1 and P4 remarked that they were difficult to interpret in situations where they were partially occluded by polygon marks due to a strong \textit{scaffold gravity}.
However, both saw this shape as novel and promising; P1 suggested using it more judiciously, such as in glyph designs comprised only of \textit{wave} marks.

Though we deliberately included a mix of asymmetric and unusual shapes in \revision{our example mark shape palette}, these comments suggest a need to exclude or reduce the likelihood that certain shapes get assigned to marks, or to preclude specific combinations of shape and encoding channel.
On the other hand, other interviewees [P3, P4, P5, P6, P8, P9] supported the \revision{idea of designer-designed mark shape palettes, comprised of shapes that could} evoke semantic associations with the underlying data or shapes that are merely \textit{`extravagant'} and \textit{`fun'} [P3].

\bstart{Adding options to the channel palette}
Of the numerical encoding channels for polygons in \revision{our example palette}, size differentiation is most salient. 
\revision{Commentary on other channels in our example palette} recalled our earlier experimentation described in \autoref{sec:reflection}, such as P1's suggestion that we could encode values into marks' stroke weights or, in reference to the \textit{star} shape, we could map values to the number of polygon vertices. 
Beyond channels that we had previously considered, P11 urged us to consider other channels, such as the fullness of a mark's fill and the inclusion of textured fill patterns.

The mark rotation channel continues to be a challenge, as our use of a circular \textit{pip} elicited some confusion. 
P2 and P3 both commented that the default position of \textit{pips} varied across mark shapes, and while marks with different shapes correspond with unique column sets whose values are not directly comparable, this is not evident at first glance.
Another recurring source of confusion was the inclusion of pips irrespective of whether the rotation channel was assigned to the mark. 
Our rationale for the universality of pips was consistency: while shapes and encoding channel combinations may differ across marks, each would exhibit this common defining characteristic, akin to how all biological diatoms have a nucleus.
\revision{P2 and P3's} comments would suggest that this uniformity conflicts with the occasional use of the pip to signify mark rotation.

\bstart{Post-sampling mark and channel overrides}
\revision{After demonstrating the} sample-based assignment of mark shapes and encoding channels, we asked \revision{our interviewees} to select a promising glyph design and tell us if and how they would refine it.
\revision{Both P10 and P11 offered suggestions relating to an} ability to override specific assignments.
P10 suggested the ability to select a mark within a single glyph design and swap out its shape or its encoding channels without affecting the rest of the design. 
He further suggested the ability to modify the output range of a selected encoding channel, such as constraining the minimum and maximum sizes of a mark or the range of possible mark fill colors. 
This hypothetical override control for a single mark suggests the need for on-demand widgets such as \textit{in-context sliders} as described by Webb~\etal~\cite{webb2008context}.
Alternatively, P11 suggested a shelf and pill interface similar to Tableau~\cite{Stolte2002Polaris}, in which  assignments to a selected mark could be overwritten by dragging alternative mark or channel pills to a shelf.

\subsection{Observations on Scaffolds and Mark Arrangements}
\label{sec:study:arranging}

\revision{We explained the concept of a glyph scaffold and a scaffold's \textit{gravity}, indicating that both were randomly determined.} 
\revision{In general, we noted that glyph designs with either extremely strong or extremely weak gravities tended to be ignored or dismissed as unpromising by interviewees; an exception was P4, who expressed an appreciation for designs with strong gravities}: \textit{``I like that all of the shapes are drawn in the center; I start seeing them as a whole, that everything has its center''}.

Opinions varied in terms of what a desirable set of options should be for a scaffold shape palette.
For instance, P9 expressed a preference for polygon scaffolds that exhibited mark placement symmetry, while
P3 and P8 expressed a preference for simple scaffold shapes over hexagons and spirals. 
The former noted that \textit{``it's easier to register the differences when the spatial organization is basic''} and the latter indicated a preference for the simplicity of a vertical linear scaffold.

\bstart{Refining the scaffolds}
As with mark and channel assignments, interviewees suggested ways to refine the scaffolds after they are assigned, such as modifying their colors [P2, P6, P11] or sizes relative to the marks superimposed on them [P5, P6, P9].
\revision{P4 and P5 both expressed an interest in reining in randomness associated with a scaffold, such as by binding its color to a value from the corresponding data point.}
Similarly, when we suggested the possibility of binding the gravity of a scaffold to a data value, so that glyphs could be differentiated by the proximity of their marks, P8 agreed that this too would be worth experimenting with.

\bstart{Mark arrangement hierarchies}
\revision{P5 and P10} asked us about the placement and ordering of marks relative to a scaffold.
Upon explaining this process (\autoref{sec:diatoms:sampling}), it was evident that both wanted to manipulate these initial placements.
\revision{P10's} suggestion was the ability to establish a visual hierarchy of marks, such that one mark is more salient than the others; this mark could be noticeably larger than the others or placed more centrally within the scaffold while other marks orbit around the scaffold's periphery.
Creating this visual hierarchy need not take place after sampling; in discussion with P5, we realized that this assignment of focal and peripheral marks could take place during column set assignment.
Once a hierarchy is defined, we could assign less salient encoding channels to peripheral marks; for instance, both P2 and P4 commented on the relative subtlety of the \textit{alpha level} channel relative to the more salient \textit{size} channel, with P4 indicating that she would relegate the former channel to marks that were less central to the visual hierarchy. 
Finally, the incorporation of a visual hierarchy could add clarity to the use of the rotation channel for peripheral marks; P10 suggested that rather than rotate these marks relative to absolute cardinal directions, they could be rotated toward or away from some other point of reference, such as a central focal mark.

\subsection{Semantic and Figurative Associations}
\label{sec:study:figurative}

Associating the visual properties of a glyph with the semantics of the underlying data or with figurative motifs was another \revision{line of questioning that we pursued}.
We suggested that some associations could be planned prior to sampling, while other associations occur serendipitously after sampling: a post-hoc recognition of emergent visual phenomena.

\bstart{Associations by design}
\revision{P6, P8, and P11} offered several examples of how \revision{designer-defined} palettes of shape and channel options could evoke aspects of the dataset. 
As a designer who has worked on map-based visualization projects, P8 suggested the use of Diatoms for generating weather-related glyphs, so that incorporating color palettes that are conventional in weather maps could result in glyphs that trigger meteorological associations.
Similarly, P11 mentioned his preference for what he referred to as the \textit{`visceral meanings'} of soft color palettes and organic shapes, particularly when drawing glyphs that represent people.
Considering the urban mobility dataset, P6 suggested incorporating the shape of a city's geographic footprint into the glyph's scaffold, so as to reinforce the association with intra-city travel.
A contrasting view came from P9, who saw planned semantic relationships as a bonus. 
Upon seeing the output of Diatoms, he remarked that \textit{``you can get a lot of inspiration from what is already here}, [although] \textit{importing shapes would work in certain situations.''} 

\bstart{Emergent associations}
On several occasions, interviewees serendipitously recognized certain shape, channel, and scaffold combinations that evoked figurative elements that aligned nicely with the underlying data.
Examining glyphs generated for the film dataset, P3 noted how certain combinations of shapes along a horizontal scaffold were reminiscent of film cameras. 
Similarly, P4 spotted a design where the length of a \textit{wave} mark corresponded with a film's runtime, evoking a physical filmstrip; in another design, a rotated \textit{circle} mark and its \textit{pip} suggested an analog clock, which was deemed appropriate for conveying a time-related value.
Some associations were less overt, such as in the context of the urban mobility dataset, where P7 spotted skyscrapers comprised of \textit{square} marks arranged in vertical scaffolds, or where 
P5 saw frequency-varying \textit{wave} marks superimposed over \textit{square} marks, a combination that evoked either the density or frenetic activity of a city. 

\lettrine[lines=9,findent=2mm,nindent=-.5mm]{\includegraphics[width=33mm]{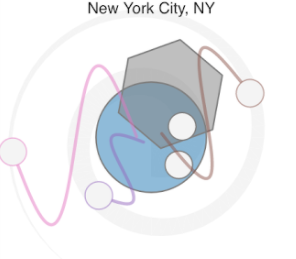}}Some emergent figurative associations are unrelated to the data. 
After using the term \textit{`personality'} to describe a particular design, P12 explained that with Diatoms, \textit{``you are creating a new entity, a new organism,''} citing the project's biological namesake.
Later in the interview, she would refer to her favourite glyph design as \textit{`Bob,'} noted for his eyes and wavy limbs (see inset at left).

\subsection{\revision{Study Limitations}}
\label{sec:study:limitations}


\revision{While our study provided us with designers' impressions of our technique and how it might be incorporated into their workflows, we cannot make any claims regarding the efficacy of the glyph designs that Diatoms ultimately inspires. 
Diatoms can provide alternative glyph designs as starting points for designers, who may in turn incorporate selective aspects of these designs into their final visualization design. 
For instance, they may iteratively adjust the glyph design and augment it with helpful annotations and legends,  while simultaneously integrating other sources of design inspiration.  
A longitudinal diary study of designers who incorporate Diatoms into their workflow could cast light on the efficacy of the technique, such as by capturing the lineage of a Diatoms-inspired glyph design.
When designers publish or disseminate their work, we can additionally assess the efficacy of their final glyph designs among their intended viewing audiences.}


\revision{Ultimately, our experience suggests that this area of research is ripe for methodological innovation, one in need of new methods and metrics for evaluating visualization design inspiration techniques.}


\section{Discussion}
\label{sec:discussion}
\revision{We see Diatoms as being part of a larger effort aimed at expanding the vocabulary of visualization design choices and combinations.
Echoing Johnson~\etal~\cite{johnson2019artifact}, we see this effort as a way to avoid converging on a local maximum, a point where most programmatically-generated visualization exhibits a common aesthetic, one with a limited potential to evoke a range of affective responses from viewers.
}

\bstart{\revision{Randomness and designer agency}}
\revision{To paraphrase P11 from our study, random sampling from palettes of marks and channels can at times result in the serendipitous discovery of a promising design; however, in many cases, it will not lead to a \textit{useful} expansion of the vocabulary for visualization design choices.
Consequently, we must temper randomness and provide designers with the agency to curate and constrain the output of a technique such as Diatoms, such as by providing the ability to cull unpromising designs or the ability to incrementally add data columns or column set designations to a glyph design specification.
Beyond assuming the role of a curator, another potential way to restore designer agency is to assign them the role of a \textit{breeder}: as inspiring precedents, we look to how Morph~\cite{morph2018} \textit{mutates} an encoding, or how House~\etal~\cite{house2006genetic} incorporated genetic algorithms into flow visualization design, and we envision similar approaches being applied to explore specific paths through the glyph design space.
}

\bstart{\revision{Designer-defined palettes}}
\revision{We see the ability to define project-specific palettes of shapes and encoding channels as a way to promote designer agency with a sample-based technique such as Diatoms.
Beyond palettes of custom polygons, paths, and colors, we envision the incorporation of external sources of visual imagery.}
For instance, DataQuilt~\cite{zhang2020dataquilt} allows visualization designers to extract mark shapes and fills from regions of photographs, while artifact-based rendering~\cite{johnson2019artifact} involves appropriating the visual texture and shape from 3D scans of small objects, such as small clay sculptures. 
These techniques could serve as a potential precursor to the palette sampling that Diatoms performs, \revision{wherein a designer provides palettes of shapes, colors, and textures extracted or scanned from images that are semantically related to the dataset at hand.} 


\bstart{\revision{From design inspiration to visualization authoring}}
Diatoms takes us closer to uniting the processes of visualization design and visualization authoring.
As existing visualization construction tools assume that people already have a design in mind prior to using them~\cite{satyanarayan2019critical}, Diatoms could inform these designs as a part of these tools.

We see different possible trajectories for integrating Diatoms into visualization construction workflows.
Our proof-of-concept implementation of Diatoms in Tableau represents one possible pipeline, from data shaping and exploratory data analysis to publishing communicative glyph-based information graphics on Tableau Public~\cite{tableaupublic}.
However, to complete this pipeline, designers would require the ability to refine, position, and format glyphs before publishing them.

Alternatively, we envision the integration of Diatoms' shape and channel sampling process into recent bespoke construction tools~\cite{satyanarayan2019critical} as an alternative to drag-and-drop data binding interactions.
Moreover, these tools could benefit from the addition of interactive design externalization and of \textit{small permutables} in particular: the ability to see the same data point visualized many different ways across a single display.
For example, this externalization could manifest as a peripheral or collapsible view akin to StructGraphics' gallery of saved templates~\cite{tsandilas2020structgraphics}.
This automated externalization could reduce the need to manually capture and arrange design artifacts in external tools.

Finally, we also foresee a use for Diatoms in multi-tool workflows~\cite{bigelow2016iterating}, whereby a standalone Diatoms implementation could export promising yet incomplete glyph designs as vector images that could be further refined using illustration tools or visualization libraries.

\section{Conclusion and Future Work}
\label{sec:conclusion}

\revision{In this paper, we introduced Diatoms, a technique for inspiring glyph design through the use of a sample-based generative process combined with interactive design externalization.
We demonstrated the technique via an extension to Tableau, which featured three example palettes to sample from: one for mark shapes, one for encoding channels, and one for glyph scaffold shapes.
We also reflected on the evolution of the technique and our experimentation with palettes that preceded those  used in the study and featured in Figures \ref{fig:teaser}--\ref{fig:dashboard}.
Finally, we collected responses to the technique from a group of information design students and professional designers, which suggested ways by which they could incorporate it into visualization design workflows.
}

Looking to the future, we hope that our research motivates others in our research community to consider the \revision{use of generative design processes and interactive design externalization} into visualization construction workflows.
\revision{We also encourage designers and researchers to experiment with mark shape, encoding channel, and scaffold shape palettes beyond those featured in this paper.}

\revision{Our interviewing of information design students led us to consider the pedagogical potential of Diatoms.
A collaborative review of glyphs generated by this technique could complement existing activities that encourage divergent thinking about visual encoding, such as sketching two quantities in as many ways as one can think of~\cite{ortiz2012}.
Diatoms could be used to introduce concepts like conjunction encoding, the relative effectiveness of integral and separable visual channels~\cite{ware2004information}, and Gestalt principles of perceptual grouping.
Beyond visual perception, the technique could also be used to explore the impedance match between the semantics of a data column and its visual representation within a glyph.
Finally, we hypothesize that an interactive \textit{small permutables} viewing mode could facilitate both types of pedagogical discussion.
}

Beyond design inspiration, we pose a speculative question that looks beyond the scope of this paper: could the Diatoms technique be used as a tool for exploratory data analysis? 
\revision{In our study, we did not ask interviewees about any new insights into the data that they realized during our brief chauffeured demonstration, as our focus was on the potential utility of the technique for inspiring early glyph design.}
\revision{We therefore leave it to future work to ask whether} the combination of a constrained random sampling of marks and channels with the ability to rapidly page between alternative glyph designs could reveal patterns in the data that were previously overlooked. 
\acknowledgments{We thank Lyn Bartram, Vidya Setlur, Maureen Stone, Mieka West, and Wesley Willett for their feedback on this work.}

\bibliographystyle{abbrv-doi-hyperref}

\bibliography{diatoms-main}

\begin{thebibliography}{10}

\bibitem{supplemental}
Diatoms: Supplemental video, 2021.
\newblock \href{https://vimeo.com/576815038}{https://vimeo.com/576815038}.

\bibitem{accurat2013}
{Accurat Studio}.
\newblock {Brain Drain}, 2013.
\newblock
  \href{https://flickr.com/photos/accurat/8423908166/}{https://flickr.com/photos/accurat/8423908166}.

\bibitem{illustrator}
{Adobe Illustrator}, 2021.
\newblock
  \href{https://adobe.com/products/illustrator.html}{https://adobe.com/products/illustrator.html}.

\bibitem{andrews2014}
R.~J. Andrews.
\newblock {Creative Routines}, 2014.
\newblock
  \href{https://infowetrust.com/project/routines}{https://infowetrust.com/project/routines}.

\bibitem{Activity2020}
{Apple Activity iOS app}, 2020.
\newblock
  \href{https://tinyurl.com/iosactivity}{https://tinyurl.com/iosactivity}.

\bibitem{aseniero2016fireflies}
B.~Aseniero, C.~Perin, M.~Eggermont, and S.~Carpendale.
\newblock Fireflies: Expressive infovis inspired by biomimicry.
\newblock In {\em IEEE VIS Arts Program Exhibit}, 2016.

\bibitem{coordinator}
A.~Aufrichtig.
\newblock co\"ordinator, 2018.
\newblock
  \href{https://spotify.github.io/coordinator/}{https://spotify.github.io/coordinator}.

\bibitem{beck2017word}
\href{https://doi.org/10.1109/TVCG.2017.2674958}{F.~Beck and D.~Weiskopf}.
\newblock \href{https://doi.org/10.1109/TVCG.2017.2674958}{Word-sized graphics
  for scientific texts}.
\newblock \href{https://doi.org/10.1109/TVCG.2017.2674958}{{\em IEEE
  Transactions on Visualization and Computer Graphics (TVCG)}},
  \href{https://doi.org/10.1109/TVCG.2017.2674958}{23(6)},
  \href{https://doi.org/10.1109/TVCG.2017.2674958}{2017}.
  \href{https://doi.org/10.1109/TVCG.2017.2674958}
{DOI: {{%
10\hspace{.1pt}\discretionary{.}{%
}{.}\hspace{.4pt}1109\discretionary{/}{%
}{/}TVCG\hspace{.1pt}\discretionary{.}{%
}{.}\hspace{.4pt}2017\hspace{.1pt}\discretionary{.}{%
}{.}\hspace{.4pt}2674958}}}


\bibitem{bigcities}
{Big Cities Health Coalition, National Association of County and City Health
  Officials}.
\newblock {Big Cities Health Inventory Data Platform}, 2019.
\newblock
  \href{http://bigcitieshealth.org/city-data}{http://bigcitieshealth.org/city-data}.

\bibitem{bigelow2016iterating}
\href{https://doi.org/10.1109/TVCG.2016.2598609}{A.~Bigelow, S.~Drucker,
  D.~Fisher, and M.~Meyer}.
\newblock \href{https://doi.org/10.1109/TVCG.2016.2598609}{Iterating between
  tools to create and edit visualizations}.
\newblock \href{https://doi.org/10.1109/TVCG.2016.2598609}{{\em IEEE
  Transactions on Visualization and Computer Graphics (Proceedings of
  InfoVis)}}, \href{https://doi.org/10.1109/TVCG.2016.2598609}{23(1)},
  \href{https://doi.org/10.1109/TVCG.2016.2598609}{2017}.
  \href{https://doi.org/10.1109/TVCG.2016.2598609}
{DOI: {{%
10\hspace{.1pt}\discretionary{.}{%
}{.}\hspace{.4pt}1109\discretionary{/}{%
}{/}TVCG\hspace{.1pt}\discretionary{.}{%
}{.}\hspace{.4pt}2016\hspace{.1pt}\discretionary{.}{%
}{.}\hspace{.4pt}2598609}}}


\bibitem{borgo2013glyph}
\href{https://doi.org/10.2312/conf/EG2013/stars/039-063}{R.~Borgo, J.~Kehrer,
  D.~H. Chung, E.~Maguire, R.~S. Laramee, H.~Hauser, M.~Ward, and M.~Chen}.
\newblock \href{https://doi.org/10.2312/conf/EG2013/stars/039-063}{Glyph-based
  visualization: Foundations, design guidelines, techniques and applications.}
\newblock \href{https://doi.org/10.2312/conf/EG2013/stars/039-063}{In {\em
  Eurographics State-of-the-Art Reports (STARs)}},
  \href{https://doi.org/10.2312/conf/EG2013/stars/039-063}{2013}.
  \href{https://doi.org/10.2312/conf/EG2013/stars/039-063}
{DOI: {{%
10\hspace{.1pt}\discretionary{.}{%
}{.}\hspace{.4pt}2312\discretionary{/}{%
}{/}conf\discretionary{/}{%
}{/}EG2013\discretionary{/}{%
}{/}stars\discretionary{/}{%
}{/}039\discretionary{%
}{-}{-}063}}}


\bibitem{borkin2013makes}
\href{https://doi.org/10.1109/TVCG.2013.234}{M.~A. Borkin, A.~A. Vo,
  Z.~Bylinskii, P.~Isola, S.~Sunkavalli, A.~Oliva, and H.~Pfister}.
\newblock \href{https://doi.org/10.1109/TVCG.2013.234}{What makes a
  visualization memorable?}
\newblock \href{https://doi.org/10.1109/TVCG.2013.234}{{\em IEEE Transactions
  on Visualization and Computer Graphics (Proceedings of InfoVis)}},
  \href{https://doi.org/10.1109/TVCG.2013.234}{19(12)},
  \href{https://doi.org/10.1109/TVCG.2013.234}{2013}.
  \href{https://doi.org/10.1109/TVCG.2013.234}
{DOI: {{%
10\hspace{.1pt}\discretionary{.}{%
}{.}\hspace{.4pt}1109\discretionary{/}{%
}{/}TVCG\hspace{.1pt}\discretionary{.}{%
}{.}\hspace{.4pt}2013\hspace{.1pt}\discretionary{.}{%
}{.}\hspace{.4pt}234}}}


\bibitem{bremer2016b}
N.~Bremer.
\newblock {My Life in Vacations}.
\newblock In {\em Data Sketches}. CRC Press, 2021.
\newblock
  \href{https://vacations.visualcinnamon.com/}{https://vacations.visualcinnamon.com}.

\bibitem{bremer2016}
N.~Bremer.
\newblock {Royal Constellations}.
\newblock In {\em Data Sketches}. CRC Press, 2021.
\newblock
  \href{https://royalconstellations.visualcinnamon.com/}{https://royalconstellations.visualcinnamon.com}.

\bibitem{buxton2000large}
\href{https://doi.org/10.1109/38.851753}{W.~Buxton, G.~Fitzmaurice,
  R.~Balakrishnan, and G.~Kurtenbach}.
\newblock \href{https://doi.org/10.1109/38.851753}{Large displays in automotive
  design}.
\newblock \href{https://doi.org/10.1109/38.851753}{{\em IEEE Computer Graphics
  and Applications (CG\&A)}}, \href{https://doi.org/10.1109/38.851753}{20(4)},
  \href{https://doi.org/10.1109/38.851753}{2000}.
  \href{https://doi.org/10.1109/38.851753}
{DOI: {{%
10\hspace{.1pt}\discretionary{.}{%
}{.}\hspace{.4pt}1109\discretionary{/}{%
}{/}38\hspace{.1pt}\discretionary{.}{%
}{.}\hspace{.4pt}851753}}}


\bibitem{byrne2019figurative}
\href{https://doi.org/10.1177/1473871617724212}{L.~Byrne, D.~Angus, and
  J.~Wiles}.
\newblock \href{https://doi.org/10.1177/1473871617724212}{Figurative frames: A
  critical vocabulary for images in information visualization}.
\newblock \href{https://doi.org/10.1177/1473871617724212}{{\em Information
  Visualization}}, \href{https://doi.org/10.1177/1473871617724212}{18(1)},
  \href{https://doi.org/10.1177/1473871617724212}{2019}.
  \href{https://doi.org/10.1177/1473871617724212}
{DOI: {{%
10\hspace{.1pt}\discretionary{.}{%
}{.}\hspace{.4pt}1177\discretionary{/}{%
}{/}1473871617724212}}}


\bibitem{chernoff1973use}
\href{https://doi.org/10.1080/01621459.1973.10482434}{H.~Chernoff}.
\newblock \href{https://doi.org/10.1080/01621459.1973.10482434}{The use of
  faces to represent points in k-dimensional space graphically}.
\newblock \href{https://doi.org/10.1080/01621459.1973.10482434}{{\em Journal of
  the American statistical Association}},
  \href{https://doi.org/10.1080/01621459.1973.10482434}{68(342)},
  \href{https://doi.org/10.1080/01621459.1973.10482434}{1973}.
  \href{https://doi.org/10.1080/01621459.1973.10482434}
{DOI: {{%
10\hspace{.1pt}\discretionary{.}{%
}{.}\hspace{.4pt}1080\discretionary{/}{%
}{/}01621459\hspace{.1pt}\discretionary{.}{%
}{.}\hspace{.4pt}1973\hspace{.1pt}\discretionary{.}{%
}{.}\hspace{.4pt}10482434}}}


\bibitem{corum2014}
J.~Corum.
\newblock {See, Think, Design Produce 2}, 2014.
\newblock Public lecture; slides \& transcript:
  \href{http://style.org/stdp2/}{http://style.org/stdp2}.

\bibitem{cruz2019dendrochronology}
\href{https://doi.org/10.1075/idj.25.1.01cru}{P.~Cruz, J.~Wihbey, A.~Ghael,
  F.~Shibuya, and S.~Costa}.
\newblock \href{https://doi.org/10.1075/idj.25.1.01cru}{Dendrochronology of
  {US} immigration}.
\newblock \href{https://doi.org/10.1075/idj.25.1.01cru}{{\em Information Design
  Journal}}, \href{https://doi.org/10.1075/idj.25.1.01cru}{25(1)},
  \href{https://doi.org/10.1075/idj.25.1.01cru}{2019}.
  \href{https://doi.org/10.1075/idj.25.1.01cru}
{DOI: {{%
10\hspace{.1pt}\discretionary{.}{%
}{.}\hspace{.4pt}1075\discretionary{/}{%
}{/}idj\hspace{.1pt}\discretionary{.}{%
}{.}\hspace{.4pt}25\hspace{.1pt}\discretionary{.}{%
}{.}\hspace{.4pt}1\hspace{.1pt}\discretionary{.}{%
}{.}\hspace{.4pt}01cru}}}


\bibitem{dvs}
{The Data Visualization Society}, 2021.
\newblock
  \href{https://datavisualizationsociety.com}{https://datavisualizationsociety.com}.

\bibitem{morph2018}
Datavized.
\newblock Morph, 2018.
\newblock
  \href{https://github.com/datavized/morph}{https://github.com/datavized/morph}.

\bibitem{davis2019design}
\href{https://doi.org/10.1075/idj.25.1.03van}{S.~B. Davis and O.~Vane}.
\newblock \href{https://doi.org/10.1075/idj.25.1.03van}{Design as
  externalization: Enabling research}.
\newblock \href{https://doi.org/10.1075/idj.25.1.03van}{{\em Information Design
  Journal}}, \href{https://doi.org/10.1075/idj.25.1.03van}{25(1)},
  \href{https://doi.org/10.1075/idj.25.1.03van}{2019}.
  \href{https://doi.org/10.1075/idj.25.1.03van}
{DOI: {{%
10\hspace{.1pt}\discretionary{.}{%
}{.}\hspace{.4pt}1075\discretionary{/}{%
}{/}idj\hspace{.1pt}\discretionary{.}{%
}{.}\hspace{.4pt}25\hspace{.1pt}\discretionary{.}{%
}{.}\hspace{.4pt}1\hspace{.1pt}\discretionary{.}{%
}{.}\hspace{.4pt}03van}}}


\bibitem{eggermont2018bio}
M.~J. Eggermont.
\newblock {\em Bio-inspired Design and Information Visualization}.
\newblock PhD thesis, University of Calgary, 2018.
\newblock
  \href{https://prism.ucalgary.ca/handle/1880/106543}{https://prism.ucalgary.ca/handle/1880/106543}.

\bibitem{eno1996}
B.~Eno.
\newblock Generative music.
\newblock {\em In Motion Magazine}, 1996.
\newblock
  \href{https://inmotionmagazine.com/eno1.html}{https://inmotionmagazine.com/eno1.html}.

\bibitem{fuchs2016systematic}
\href{https://doi.org/10.1109/TVCG.2016.2549018}{J.~Fuchs, P.~Isenberg,
  A.~Bezerianos, and D.~Keim}.
\newblock \href{https://doi.org/10.1109/TVCG.2016.2549018}{A systematic review
  of experimental studies on data glyphs}.
\newblock \href{https://doi.org/10.1109/TVCG.2016.2549018}{{\em IEEE
  Transactions on Visualization and Computer Graphics (TVCG)}},
  \href{https://doi.org/10.1109/TVCG.2016.2549018}{23(7)},
  \href{https://doi.org/10.1109/TVCG.2016.2549018}{2016}.
  \href{https://doi.org/10.1109/TVCG.2016.2549018}
{DOI: {{%
10\hspace{.1pt}\discretionary{.}{%
}{.}\hspace{.4pt}1109\discretionary{/}{%
}{/}TVCG\hspace{.1pt}\discretionary{.}{%
}{.}\hspace{.4pt}2016\hspace{.1pt}\discretionary{.}{%
}{.}\hspace{.4pt}2549018}}}


\bibitem{gapminder}
{Gapminder}.
\newblock {Free data from World Bank via gapminder.org}, 2021.
\newblock \ccby~\href{https://gapminder.org/data/}{https://gapminder.org/data}.

\bibitem{goffin2014exploring}
\href{https://doi.org/10.1109/TVCG.2014.2346435}{P.~Goffin, W.~Willett, J.-D.
  Fekete, and P.~Isenberg}.
\newblock \href{https://doi.org/10.1109/TVCG.2014.2346435}{Exploring the
  placement and design of word-scale visualizations}.
\newblock \href{https://doi.org/10.1109/TVCG.2014.2346435}{{\em IEEE
  Transactions on Visualization and Computer Graphics (Proceedings of
  InfoVis)}}, \href{https://doi.org/10.1109/TVCG.2014.2346435}{20(12)},
  \href{https://doi.org/10.1109/TVCG.2014.2346435}{2014}.
  \href{https://doi.org/10.1109/TVCG.2014.2346435}
{DOI: {{%
10\hspace{.1pt}\discretionary{.}{%
}{.}\hspace{.4pt}1109\discretionary{/}{%
}{/}TVCG\hspace{.1pt}\discretionary{.}{%
}{.}\hspace{.4pt}2014\hspace{.1pt}\discretionary{.}{%
}{.}\hspace{.4pt}2346435}}}


\bibitem{goldsteinspaceanalysis}
\href{https://doi.org/10.22360/simaud.2017.simaud.007}{R.~Goldstein,
  S.~Breslav, K.~Walmsley, and A.~Khan}.
\newblock
  \href{https://doi.org/10.22360/simaud.2017.simaud.007}{{SpaceAnalysis}: A
  tool for pathfinding, visibility, and acoustics analyses in generative design
  workflows}.
\newblock \href{https://doi.org/10.22360/simaud.2017.simaud.007}{In {\em
  Proceedings of the Symposium on Simulation for Architecture and Urban Design
  (SIMAUD)}}, \href{https://doi.org/10.22360/simaud.2017.simaud.007}{2017}.
  \href{https://doi.org/10.22360/simaud.2017.simaud.007}
{DOI: {{%
10\hspace{.1pt}\discretionary{.}{%
}{.}\hspace{.4pt}22360\discretionary{/}{%
}{/}simaud\hspace{.1pt}\discretionary{.}{%
}{.}\hspace{.4pt}2017\hspace{.1pt}\discretionary{.}{%
}{.}\hspace{.4pt}simaud\hspace{.1pt}\discretionary{.}{%
}{.}\hspace{.4pt}007}}}


\bibitem{grammel2013survey}
\href{https://doi.org/10.2312/PE.EuroVisShort.EuroVisShort2013.019-023}{L.~Grammel,
  C.~Bennett, M.~Tory, and M.-A.~D. Storey}.
\newblock
  \href{https://doi.org/10.2312/PE.EuroVisShort.EuroVisShort2013.019-023}{A
  survey of visualization construction user interfaces}.
\newblock
  \href{https://doi.org/10.2312/PE.EuroVisShort.EuroVisShort2013.019-023}{In
  {\em Short Paper Proceedings of the Eurographics Conference on Visualization
  (EuroVis)}},
  \href{https://doi.org/10.2312/PE.EuroVisShort.EuroVisShort2013.019-023}{2013}.
  \href{https://doi.org/10.2312/PE.EuroVisShort.EuroVisShort2013.019-023}
{DOI: {{%
10\hspace{.1pt}\discretionary{.}{%
}{.}\hspace{.4pt}2312\discretionary{/}{%
}{/}PE\hspace{.1pt}\discretionary{.}{%
}{.}\hspace{.4pt}EuroVisShort\hspace{.1pt}\discretionary{.}{%
}{.}\hspace{.4pt}EuroVisShort2013\hspace{.1pt}\discretionary{.}{%
}{.}\hspace{.4pt}019\discretionary{%
}{-}{-}023}}}


\bibitem{gross2018generative}
B.~Gro{\ss}, H.~Bohnacker, J.~Laub, and C.~Lazzeroni.
\newblock {\em Generative Design: Visualize, Program, and Create with
  Javascript in {p5.js}}.
\newblock Princeton Architectural Press, 2018.
\newblock
  \href{http://www.generative-gestaltung.de/2/}{http://www.generative-gestaltung.de/2}.

\bibitem{heer2008graphical}
\href{https://doi.org/10.1109/TVCG.2008.137}{J.~Heer, J.~Mackinlay, C.~Stolte,
  and M.~Agrawala}.
\newblock \href{https://doi.org/10.1109/TVCG.2008.137}{Graphical histories for
  visualization: Supporting analysis, communication, and evaluation}.
\newblock \href{https://doi.org/10.1109/TVCG.2008.137}{{\em IEEE Transactions
  on Visualization and Computer Graphics (Proceedings of InfoVis)}},
  \href{https://doi.org/10.1109/TVCG.2008.137}{14(6)},
  \href{https://doi.org/10.1109/TVCG.2008.137}{2008}.
  \href{https://doi.org/10.1109/TVCG.2008.137}
{DOI: {{%
10\hspace{.1pt}\discretionary{.}{%
}{.}\hspace{.4pt}1109\discretionary{/}{%
}{/}TVCG\hspace{.1pt}\discretionary{.}{%
}{.}\hspace{.4pt}2008\hspace{.1pt}\discretionary{.}{%
}{.}\hspace{.4pt}137}}}


\bibitem{horn1998metaphor}
W.~Horn, C.~Popow, and L.~Unterasinger.
\newblock Metaphor graphics to visualize icu data over time.
\newblock {\em Intelligent Data Analysis in Medicine and Pharmacology
  (IDAMAP)}, 1998.
\newblock \href{https://tinyurl.com/horn2018}{https://tinyurl.com/horn2018}.

\bibitem{house2006genetic}
\href{https://doi.org/10.1109/TVCG.2006.58}{D.~H. House, A.~S. Bair, and
  C.~Ware}.
\newblock \href{https://doi.org/10.1109/TVCG.2006.58}{\revision{An approach to
  the perceptual optimization of complex visualizations}}.
\newblock \href{https://doi.org/10.1109/TVCG.2006.58}{{\em IEEE Transactions on
  Visualization and Computer Graphics (TVCG)}},
  \href{https://doi.org/10.1109/TVCG.2006.58}{12(4)},
  \href{https://doi.org/10.1109/TVCG.2006.58}{2006}.
  \href{https://doi.org/10.1109/TVCG.2006.58}
{DOI: {{%
10\hspace{.1pt}\discretionary{.}{%
}{.}\hspace{.4pt}1109\discretionary{/}{%
}{/}TVCG\hspace{.1pt}\discretionary{.}{%
}{.}\hspace{.4pt}2006\hspace{.1pt}\discretionary{.}{%
}{.}\hspace{.4pt}58}}}


\bibitem{datastories}
M.~Ignac, E.~Bertini, and M.~Stefaner.
\newblock {Data Art and Visual Programming with Marcin Ignac from Variable},
  2019.
\newblock Data Stories podcast episode \#153:
  \href{https://tinyurl.com/datastories153}{https://tinyurl.com/datastories153}.

\bibitem{imdb}
{IMDb Datasets}, 2021.
\newblock \href{https://imdb.com/interfaces/}{https://imdb.com/interfaces}.

\bibitem{BlockBuilder}
I.~Johnson.
\newblock Bl.ock builder, 2015.
\newblock \href{https://blockbuilder.org}{https://blockbuilder.org}.

\bibitem{johnson2019artifact}
\href{https://doi.org/10.1109/TVCG.2019.2934260}{S.~Johnson, F.~Samsel,
  G.~Abram, D.~Olson, A.~J. Solis, B.~Herman, P.~J. Wolfram, C.~Lenglet, and
  D.~F. Keefe}.
\newblock \href{https://doi.org/10.1109/TVCG.2019.2934260}{Artifact-based
  rendering: Harnessing natural and traditional visual media for more
  expressive and engaging {3D} visualizations}.
\newblock \href{https://doi.org/10.1109/TVCG.2019.2934260}{{\em IEEE
  Transactions on Visualization and Computer Graphics (Proceedings of
  SciVis)}}, \href{https://doi.org/10.1109/TVCG.2019.2934260}{26(1)},
  \href{https://doi.org/10.1109/TVCG.2019.2934260}{2019}.
  \href{https://doi.org/10.1109/TVCG.2019.2934260}
{DOI: {{%
10\hspace{.1pt}\discretionary{.}{%
}{.}\hspace{.4pt}1109\discretionary{/}{%
}{/}TVCG\hspace{.1pt}\discretionary{.}{%
}{.}\hspace{.4pt}2019\hspace{.1pt}\discretionary{.}{%
}{.}\hspace{.4pt}2934260}}}


\bibitem{jones2020}
A.~Jones.
\newblock {Take My Breath Away}, 2020.
\newblock Tableau Public.
  \href{https://tinyurl.com/jonestc2020}{https://tinyurl.com/jonestc2020}.

\bibitem{jhunja2021}
R.~Juhnja.
\newblock Putting three meals on the table.
\newblock {\em Nightingale}, 2021.
\newblock
  \href{https://tinyurl.com/jhunja2021}{https://tinyurl.com/jhunja2021}.

\bibitem{kim2016data}
\href{https://doi.org/10.1109/TVCG.2016.2598620}{N.~W. Kim, E.~Schweickart,
  Z.~Liu, M.~Dontcheva, W.~Li, J.~Popovic, and H.~Pfister}.
\newblock \href{https://doi.org/10.1109/TVCG.2016.2598620}{Data-driven guides:
  Supporting expressive design for information graphics}.
\newblock \href{https://doi.org/10.1109/TVCG.2016.2598620}{{\em IEEE
  Transactions on Visualization and Computer Graphics (Proceedings of
  InfoVis)}}, \href{https://doi.org/10.1109/TVCG.2016.2598620}{23(1)},
  \href{https://doi.org/10.1109/TVCG.2016.2598620}{2017}.
  \href{https://doi.org/10.1109/TVCG.2016.2598620}
{DOI: {{%
10\hspace{.1pt}\discretionary{.}{%
}{.}\hspace{.4pt}1109\discretionary{/}{%
}{/}TVCG\hspace{.1pt}\discretionary{.}{%
}{.}\hspace{.4pt}2016\hspace{.1pt}\discretionary{.}{%
}{.}\hspace{.4pt}2598620}}}


\bibitem{kleiberg2001botanical}
\href{https://doi.org/10.1109/INFVIS.2001.963285}{E.~Kleiberg, H.~van
  De~Wetering, and J.~J. van Wijk}.
\newblock \href{https://doi.org/10.1109/INFVIS.2001.963285}{Botanical
  visualization of huge hierarchies}.
\newblock \href{https://doi.org/10.1109/INFVIS.2001.963285}{In {\em IEEE
  Symposium on Information Visualization (InfoVis)}},
  \href{https://doi.org/10.1109/INFVIS.2001.963285}{2001}.
  \href{https://doi.org/10.1109/INFVIS.2001.963285}
{DOI: {{%
10\hspace{.1pt}\discretionary{.}{%
}{.}\hspace{.4pt}1109\discretionary{/}{%
}{/}INFVIS\hspace{.1pt}\discretionary{.}{%
}{.}\hspace{.4pt}2001\hspace{.1pt}\discretionary{.}{%
}{.}\hspace{.4pt}963285}}}


\bibitem{kosara2016presentation}
\href{https://doi.org/10.1109/MCG.2016.2}{R.~Kosara}.
\newblock \href{https://doi.org/10.1109/MCG.2016.2}{Presentation-oriented
  visualization techniques}.
\newblock \href{https://doi.org/10.1109/MCG.2016.2}{{\em IEEE Computer Graphics
  and Applications (CG\&A)}}, \href{https://doi.org/10.1109/MCG.2016.2}{36(1)},
  \href{https://doi.org/10.1109/MCG.2016.2}{2016}.
  \href{https://doi.org/10.1109/MCG.2016.2}
{DOI: {{%
10\hspace{.1pt}\discretionary{.}{%
}{.}\hspace{.4pt}1109\discretionary{/}{%
}{/}MCG\hspace{.1pt}\discretionary{.}{%
}{.}\hspace{.4pt}2016\hspace{.1pt}\discretionary{.}{%
}{.}\hspace{.4pt}2}}}


\bibitem{kovacs2018}
I.~Kov\'acs.
\newblock {Gender \& Ethnic Disparities in Tech Companies}, 2018.
\newblock Tableau Public.
  \href{https://tinyurl.com/kovacs2018}{https://tinyurl.com/kovacs2018}.

\bibitem{koyama2020sequential}
\href{https://doi.org/10.1145/3386569.3392444}{Y.~Koyama, I.~Sato, and
  M.~Goto}.
\newblock \href{https://doi.org/10.1145/3386569.3392444}{Sequential gallery for
  interactive visual design optimization}.
\newblock \href{https://doi.org/10.1145/3386569.3392444}{{\em ACM Transactions
  on Graphics (Proceedings of SIGGRAPH)}},
  \href{https://doi.org/10.1145/3386569.3392444}{39(4)},
  \href{https://doi.org/10.1145/3386569.3392444}{2020}.
  \href{https://doi.org/10.1145/3386569.3392444}
{DOI: {{%
10\hspace{.1pt}\discretionary{.}{%
}{.}\hspace{.4pt}1145\discretionary{/}{%
}{/}3386569\hspace{.1pt}\discretionary{.}{%
}{.}\hspace{.4pt}3392444}}}


\bibitem{xenographics}
M.~Lambrechts.
\newblock Xenographics: Weird but (sometimes) useful charts, 2021.
\newblock \href{https://xeno.graphics/}{https://xeno.graphics}.

\bibitem{lee2010designing}
\href{https://doi.org/10.1145/1753326.1753667}{B.~Lee, S.~Srivastava, R.~Kumar,
  R.~Brafman, and S.~R. Klemmer}.
\newblock \href{https://doi.org/10.1145/1753326.1753667}{Designing with
  interactive example galleries}.
\newblock \href{https://doi.org/10.1145/1753326.1753667}{In {\em Proceedings of
  the ACM Conference on Human Factors in Computing Systems (CHI)}},
  \href{https://doi.org/10.1145/1753326.1753667}{2010}.
  \href{https://doi.org/10.1145/1753326.1753667}
{DOI: {{%
10\hspace{.1pt}\discretionary{.}{%
}{.}\hspace{.4pt}1145\discretionary{/}{%
}{/}1753326\hspace{.1pt}\discretionary{.}{%
}{.}\hspace{.4pt}1753667}}}


\bibitem{liu2018data}
\href{https://doi.org/10.1145/3173574.3173697}{Z.~Liu, J.~Thompson, A.~Wilson,
  M.~Dontcheva, J.~Delorey, S.~Grigg, B.~Kerr, and J.~Stasko}.
\newblock \href{https://doi.org/10.1145/3173574.3173697}{{Data Illustrator}:
  Augmenting vector design tools with lazy data binding for expressive
  visualization authoring}.
\newblock \href{https://doi.org/10.1145/3173574.3173697}{In {\em Proceedings of
  the ACM Conference on Human Factors in Computing Systems (CHI)}},
  \href{https://doi.org/10.1145/3173574.3173697}{2018}.
  \href{https://doi.org/10.1145/3173574.3173697}
{DOI: {{%
10\hspace{.1pt}\discretionary{.}{%
}{.}\hspace{.4pt}1145\discretionary{/}{%
}{/}3173574\hspace{.1pt}\discretionary{.}{%
}{.}\hspace{.4pt}3173697}}}


\bibitem{lloyd2011human}
\href{https://doi.org/10.1109/TVCG.2011.209}{D.~Lloyd and J.~Dykes}.
\newblock \href{https://doi.org/10.1109/TVCG.2011.209}{Human-centered
  approaches in geovisualization design: Investigating multiple methods through
  a long-term case study}.
\newblock \href{https://doi.org/10.1109/TVCG.2011.209}{{\em IEEE Transactions
  on Visualization and Computer Graphics (Proceedings of InfoVis)}},
  \href{https://doi.org/10.1109/TVCG.2011.209}{17(12)},
  \href{https://doi.org/10.1109/TVCG.2011.209}{2011}.
  \href{https://doi.org/10.1109/TVCG.2011.209}
{DOI: {{%
10\hspace{.1pt}\discretionary{.}{%
}{.}\hspace{.4pt}1109\discretionary{/}{%
}{/}TVCG\hspace{.1pt}\discretionary{.}{%
}{.}\hspace{.4pt}2011\hspace{.1pt}\discretionary{.}{%
}{.}\hspace{.4pt}209}}}


\bibitem{Lupi2016DearData}
G.~Lupi and S.~Posavec.
\newblock {\em Dear Data}.
\newblock Princeton Architectural Press, 2016.

\bibitem{Lupi2018OCD}
G.~Lupi and S.~Posavec.
\newblock {\em Observe, Collect, Draw!: A Visual Journal}.
\newblock Princeton Architectural Press, 2018.

\bibitem{mackinlay2007showme}
\href{https://doi.org/10.1109/TVCG.2007.70594}{J.~Mackinlay, P.~Hanrahan, and
  C.~Stolte}.
\newblock \href{https://doi.org/10.1109/TVCG.2007.70594}{{Show Me}: Automatic
  presentation for visual analysis}.
\newblock \href{https://doi.org/10.1109/TVCG.2007.70594}{{\em IEEE Transactions
  on Visualization and Computer Graphics (TVCG)}},
  \href{https://doi.org/10.1109/TVCG.2007.70594}{13(6)},
  \href{https://doi.org/10.1109/TVCG.2007.70594}{2007}.
  \href{https://doi.org/10.1109/TVCG.2007.70594}
{DOI: {{%
10\hspace{.1pt}\discretionary{.}{%
}{.}\hspace{.4pt}1109\discretionary{/}{%
}{/}TVCG\hspace{.1pt}\discretionary{.}{%
}{.}\hspace{.4pt}2007\hspace{.1pt}\discretionary{.}{%
}{.}\hspace{.4pt}70594}}}


\bibitem{maguire2012taxonomy}
\href{https://doi.org/10.1109/TVCG.2012.271}{E.~Maguire, P.~Rocca-Serra, S.-A.
  Sansone, J.~Davies, and M.~Chen}.
\newblock \href{https://doi.org/10.1109/TVCG.2012.271}{Taxonomy-based glyph
  design—with a case study on visualizing workflows of biological
  experiments}.
\newblock \href{https://doi.org/10.1109/TVCG.2012.271}{{\em IEEE Transactions
  on Visualization and Computer Graphics (Proceedings of InfoVis)}},
  \href{https://doi.org/10.1109/TVCG.2012.271}{18(12)},
  \href{https://doi.org/10.1109/TVCG.2012.271}{2012}.
  \href{https://doi.org/10.1109/TVCG.2012.271}
{DOI: {{%
10\hspace{.1pt}\discretionary{.}{%
}{.}\hspace{.4pt}1109\discretionary{/}{%
}{/}TVCG\hspace{.1pt}\discretionary{.}{%
}{.}\hspace{.4pt}2012\hspace{.1pt}\discretionary{.}{%
}{.}\hspace{.4pt}271}}}


\bibitem{marks1997design}
\href{https://doi.org/10.1145/258734.258887}{J.~Marks, B.~Andalman, P.~A.
  Beardsley, W.~Freeman, S.~Gibson, J.~Hodgins, T.~Kang, B.~Mirtich,
  H.~Pfister, W.~Ruml, et~al.}
\newblock \href{https://doi.org/10.1145/258734.258887}{Design galleries: A
  general approach to setting parameters for computer graphics and animation}.
\newblock \href{https://doi.org/10.1145/258734.258887}{In {\em Proceedings of
  the ACM Conference on Computer Graphics and Interactive Techniques
  (SIGGRAPH)}}, \href{https://doi.org/10.1145/258734.258887}{1997}.
  \href{https://doi.org/10.1145/258734.258887}
{DOI: {{%
10\hspace{.1pt}\discretionary{.}{%
}{.}\hspace{.4pt}1145\discretionary{/}{%
}{/}258734\hspace{.1pt}\discretionary{.}{%
}{.}\hspace{.4pt}258887}}}


\bibitem{matejka2018dream}
\href{https://doi.org/10.1145/3173574.3173943}{J.~Matejka, M.~Glueck,
  E.~Bradner, A.~Hashemi, T.~Grossman, and G.~Fitzmaurice}.
\newblock \href{https://doi.org/10.1145/3173574.3173943}{Dream lens:
  Exploration and visualization of large-scale generative design datasets}.
\newblock \href{https://doi.org/10.1145/3173574.3173943}{In {\em Proceedings of
  the ACM Conference on Human Factors in Computing Systems (CHI)}},
  \href{https://doi.org/10.1145/3173574.3173943}{2018}.
  \href{https://doi.org/10.1145/3173574.3173943}
{DOI: {{%
10\hspace{.1pt}\discretionary{.}{%
}{.}\hspace{.4pt}1145\discretionary{/}{%
}{/}3173574\hspace{.1pt}\discretionary{.}{%
}{.}\hspace{.4pt}3173943}}}


\bibitem{okuya2020investigating}
\href{https://doi.org/10.1145/3313831.3376736}{Y.~Okuya, O.~Gladin,
  N.~Ladev{\`e}ze, C.~Fleury, and P.~Bourdot}.
\newblock \href{https://doi.org/10.1145/3313831.3376736}{Investigating
  collaborative exploration of design alternatives on a wall-sized display}.
\newblock \href{https://doi.org/10.1145/3313831.3376736}{In {\em Proceedings of
  the ACM Conference on Human Factors in Computing Systems (CHI)}},
  \href{https://doi.org/10.1145/3313831.3376736}{2020}.
  \href{https://doi.org/10.1145/3313831.3376736}
{DOI: {{%
10\hspace{.1pt}\discretionary{.}{%
}{.}\hspace{.4pt}1145\discretionary{/}{%
}{/}3313831\hspace{.1pt}\discretionary{.}{%
}{.}\hspace{.4pt}3376736}}}


\bibitem{ortiz2012}
S.~Ortiz.
\newblock 45 ways to communicate two quantities.
\newblock {\em visual.ly}, 2012.
\newblock
  \href{http://blog.visual.ly/45-ways-to-communicate-two-quantities/}{http://blog.visual.ly/45-ways-to-communicate-two-quantities/}.

\bibitem{parsons2020}
\href{https://doi.org/10.1109/VIS47514.2020.00042}{P.~Parsons, C.~M. Gray,
  A.~Baigelenov, and I.~Carr}.
\newblock \href{https://doi.org/10.1109/VIS47514.2020.00042}{Design judgment in
  data visualization practice}.
\newblock \href{https://doi.org/10.1109/VIS47514.2020.00042}{In {\em In Short
  Paper Proceedings of the IEEE Conference on Visualization (VIS)}},
  \href{https://doi.org/10.1109/VIS47514.2020.00042}{2020}.
  \href{https://doi.org/10.1109/VIS47514.2020.00042}
{DOI: {{%
10\hspace{.1pt}\discretionary{.}{%
}{.}\hspace{.4pt}1109\discretionary{/}{%
}{/}VIS47514\hspace{.1pt}\discretionary{.}{%
}{.}\hspace{.4pt}2020\hspace{.1pt}\discretionary{.}{%
}{.}\hspace{.4pt}00042}}}


\bibitem{perin2014revisiting}
\href{https://doi.org/10.1109/TVCG.2014.2346279}{C.~Perin, P.~Dragicevic, and
  J.-D. Fekete}.
\newblock \href{https://doi.org/10.1109/TVCG.2014.2346279}{Revisiting {Bertin}
  matrices: New interactions for crafting tabular visualizations}.
\newblock \href{https://doi.org/10.1109/TVCG.2014.2346279}{{\em IEEE
  Transactions on Visualization and Computer Graphics (Proceedings of
  InfoVis)}}, \href{https://doi.org/10.1109/TVCG.2014.2346279}{20(12)},
  \href{https://doi.org/10.1109/TVCG.2014.2346279}{2014}.
  \href{https://doi.org/10.1109/TVCG.2014.2346279}
{DOI: {{%
10\hspace{.1pt}\discretionary{.}{%
}{.}\hspace{.4pt}1109\discretionary{/}{%
}{/}TVCG\hspace{.1pt}\discretionary{.}{%
}{.}\hspace{.4pt}2014\hspace{.1pt}\discretionary{.}{%
}{.}\hspace{.4pt}2346279}}}


\bibitem{ren2018charticulator}
\href{https://doi.org/10.1109/TVCG.2018.2865158}{D.~Ren, B.~Lee, and
  M.~Brehmer}.
\newblock \href{https://doi.org/10.1109/TVCG.2018.2865158}{Charticulator:
  Interactive construction of bespoke chart layouts}.
\newblock \href{https://doi.org/10.1109/TVCG.2018.2865158}{{\em IEEE
  Transactions on Visualization and Computer Graphics (Proceedings of
  InfoVis)}}, \href{https://doi.org/10.1109/TVCG.2018.2865158}{25(1)},
  \href{https://doi.org/10.1109/TVCG.2018.2865158}{2019}.
  \href{https://doi.org/10.1109/TVCG.2018.2865158}
{DOI: {{%
10\hspace{.1pt}\discretionary{.}{%
}{.}\hspace{.4pt}1109\discretionary{/}{%
}{/}TVCG\hspace{.1pt}\discretionary{.}{%
}{.}\hspace{.4pt}2018\hspace{.1pt}\discretionary{.}{%
}{.}\hspace{.4pt}2865158}}}


\bibitem{ren2018reflecting}
D.~Ren, B.~Lee, M.~Brehmer, and N.~Henry~Riche.
\newblock Reflecting on the evaluation of visualization authoring systems.
\newblock In {\em Workshop Proceedings of Evaluation and Beyond -
  Methodological Approaches for Visualization (BELIV)}, 2018.
\newblock \href{https://aka.ms/renbeliv18}{https://aka.ms/renbeliv18}.

\bibitem{Blocksplorer}
I.~Ros.
\newblock bl.ocksplorer, 2015.
\newblock \href{https://bl.ocksplorer.org}{https://bl.ocksplorer.org}.

\bibitem{diatoms:wikimedia}
Rovag.
\newblock Diatoms, 2009.
\newblock \ccby~3.0 via Wikimedia Commons:
  \href{https://commons.wikimedia.org/wiki/File:Diatoms.jpg}{https://commons.wikimedia.org/wiki/File:Diatoms.jpg}.

\bibitem{ruzicka2020}
E.~Ruzicka.
\newblock Socializing in a world of social distance: A {COVID-19} data journey.
\newblock {\em Nightingale}, 2020.
\newblock
  \href{https://tinyurl.com/ruzicka2020}{https://tinyurl.com/ruzicka2020}.

\bibitem{satyanarayan2014lyra}
\href{https://doi.org/10.1111/cgf.12391}{A.~Satyanarayan and J.~Heer}.
\newblock \href{https://doi.org/10.1111/cgf.12391}{Lyra: An interactive
  visualization design environment}.
\newblock \href{https://doi.org/10.1111/cgf.12391}{{\em Computer Graphics Forum
  (Proceedings of EuroVis)}}, \href{https://doi.org/10.1111/cgf.12391}{33(3)},
  \href{https://doi.org/10.1111/cgf.12391}{2014}.
  \href{https://doi.org/10.1111/cgf.12391}
{DOI: {{%
10\hspace{.1pt}\discretionary{.}{%
}{.}\hspace{.4pt}1111\discretionary{/}{%
}{/}cgf\hspace{.1pt}\discretionary{.}{%
}{.}\hspace{.4pt}12391}}}


\bibitem{satyanarayan2019critical}
\href{https://doi.org/10.1109/TVCG.2019.2934281}{A.~Satyanarayan, B.~Lee,
  D.~Ren, J.~Heer, J.~Stasko, J.~Thompson, M.~Brehmer, and Z.~Liu}.
\newblock \href{https://doi.org/10.1109/TVCG.2019.2934281}{Critical reflections
  on visualization authoring systems}.
\newblock \href{https://doi.org/10.1109/TVCG.2019.2934281}{{\em IEEE
  Transactions on Visualization and Computer Graphics (Proceedings of
  InfoVis)}}, \href{https://doi.org/10.1109/TVCG.2019.2934281}{26(1)},
  \href{https://doi.org/10.1109/TVCG.2019.2934281}{2020}.
  \href{https://doi.org/10.1109/TVCG.2019.2934281}
{DOI: {{%
10\hspace{.1pt}\discretionary{.}{%
}{.}\hspace{.4pt}1109\discretionary{/}{%
}{/}TVCG\hspace{.1pt}\discretionary{.}{%
}{.}\hspace{.4pt}2019\hspace{.1pt}\discretionary{.}{%
}{.}\hspace{.4pt}2934281}}}


\bibitem{procreate}
{Savage Interactive Pty Ltd}.
\newblock Procreate, 2019.
\newblock \href{https://procreate.art/ipad}{https://procreate.art/ipad}.

\bibitem{schroeder2015visualization}
\href{https://doi.org/10.1109/TVCG.2015.2467153}{D.~Schroeder and D.~F. Keefe}.
\newblock
  \href{https://doi.org/10.1109/TVCG.2015.2467153}{\revision{Visualization-by-sketching:
  An artist's interface for creating multivariate time-varying data
  visualizations}}.
\newblock \href{https://doi.org/10.1109/TVCG.2015.2467153}{{\em IEEE
  Transactions on Visualization and Computer Graphics (Proceedings of
  SciVis)}}, \href{https://doi.org/10.1109/TVCG.2015.2467153}{22(1)},
  \href{https://doi.org/10.1109/TVCG.2015.2467153}{2016}.
  \href{https://doi.org/10.1109/TVCG.2015.2467153}
{DOI: {{%
10\hspace{.1pt}\discretionary{.}{%
}{.}\hspace{.4pt}1109\discretionary{/}{%
}{/}TVCG\hspace{.1pt}\discretionary{.}{%
}{.}\hspace{.4pt}2015\hspace{.1pt}\discretionary{.}{%
}{.}\hspace{.4pt}2467153}}}


\bibitem{SchulteKovacs2019}
K.~Schulte and I.~Kov\'acs.
\newblock ``{Surprise Me}'': Creating advanced and unique charts, 2019.
\newblock Tableau Conference presentation; video:
  \href{https://youtu.be/AZ3tGb5t8KE}{https://youtu.be/AZ3tGb5t8KE}.

\bibitem{delve}
{Sidewalk Labs}.
\newblock {Delve}, 2021.
\newblock
  \href{https://hello.delve.sidewalklabs.com/}{https://hello.delve.sidewalklabs.com}.

\bibitem{smart2019measuring}
\href{https://doi.org/10.1145/3290605.3300899}{S.~Smart and D.~A. Szafir}.
\newblock \href{https://doi.org/10.1145/3290605.3300899}{\revision{Measuring
  the separability of shape, size, and color in scatterplots}}.
\newblock \href{https://doi.org/10.1145/3290605.3300899}{In {\em Proceedings of
  the ACM Conference on Human Factors in Computing Systems (CHI)}},
  \href{https://doi.org/10.1145/3290605.3300899}{2019}.
  \href{https://doi.org/10.1145/3290605.3300899}
{DOI: {{%
10\hspace{.1pt}\discretionary{.}{%
}{.}\hspace{.4pt}1145\discretionary{/}{%
}{/}3290605\hspace{.1pt}\discretionary{.}{%
}{.}\hspace{.4pt}3300899}}}


\bibitem{stefaner2014}
M.~Stefaner and D.~Baur.
\newblock {OECD Regional Well-Being}, 2014.
\newblock
  \href{https://truth-and-beauty.net/projects/oecd-regional-wellbeing/}{https://truth-and-beauty.net/projects/oecd-regional-wellbeing}.

\bibitem{stefaner2017}
M.~Stefaner, F.~Rausch, J.~Leist, M.~Paeschke, D.~Baur, and T.~Kekeritz.
\newblock {OECD Better Life Index}, 2017.
\newblock
  \href{https://truth-and-beauty.net/projects/oecd-better-life-index}{https://truth-and-beauty.net/projects/oecd-better-life-index}.

\bibitem{Stolte2002Polaris}
\href{https://doi.org/10.1109/2945.981851}{C.~Stolte, D.~Tang, and
  P.~Hanrahan}.
\newblock \href{https://doi.org/10.1109/2945.981851}{Polaris: A system for
  query, analysis, and visualization of multidimensional relational databases}.
\newblock \href{https://doi.org/10.1109/2945.981851}{{\em IEEE Transactions on
  Visualization and Computer Graphics (TVCG)}},
  \href{https://doi.org/10.1109/2945.981851}{8(1)},
  \href{https://doi.org/10.1109/2945.981851}{2002}.
  \href{https://doi.org/10.1109/2945.981851}
{DOI: {{%
10\hspace{.1pt}\discretionary{.}{%
}{.}\hspace{.4pt}1109\discretionary{/}{%
}{/}2945\hspace{.1pt}\discretionary{.}{%
}{.}\hspace{.4pt}981851}}}


\bibitem{swearngin2020scout}
\href{https://doi.org/10.1145/3313831.3376593}{A.~Swearngin, C.~Wang,
  A.~Oleson, J.~Fogarty, and A.~J. Ko}.
\newblock \href{https://doi.org/10.1145/3313831.3376593}{Scout: Rapid
  exploration of interface layout alternatives through high-level design
  constraints}.
\newblock \href{https://doi.org/10.1145/3313831.3376593}{In {\em Proceedings of
  the ACM Conference on Human Factors in Computing Systems (CHI)}},
  \href{https://doi.org/10.1145/3313831.3376593}{2020}.
  \href{https://doi.org/10.1145/3313831.3376593}
{DOI: {{%
10\hspace{.1pt}\discretionary{.}{%
}{.}\hspace{.4pt}1145\discretionary{/}{%
}{/}3313831\hspace{.1pt}\discretionary{.}{%
}{.}\hspace{.4pt}3376593}}}


\bibitem{tableaupublic}
{Tableau Public}, 2021.
\newblock \href{https://public.tableau.com/s/}{https://public.tableau.com/s}.

\bibitem{tableau}
{Tableau Desktop}, 2021.
\newblock
  \href{https://www.tableau.com/products/desktop}{https://www.tableau.com/products/desktop}.

\bibitem{tableauext}
{Tableau Extensions API}, 2021.
\newblock
  \href{https://tableau.com/developer/extensions}{https://tableau.com/developer/extensions}.

\bibitem{terry2002side}
\href{https://doi.org/10.1145/571985.571996}{M.~Terry and E.~D. Mynatt}.
\newblock \href{https://doi.org/10.1145/571985.571996}{Side views: persistent,
  on-demand previews for open-ended tasks}.
\newblock \href{https://doi.org/10.1145/571985.571996}{In {\em Proceedings of
  the ACM Symposium on User Interface Software and Technology (UIST)}},
  \href{https://doi.org/10.1145/571985.571996}{2002}.
  \href{https://doi.org/10.1145/571985.571996}
{DOI: {{%
10\hspace{.1pt}\discretionary{.}{%
}{.}\hspace{.4pt}1145\discretionary{/}{%
}{/}571985\hspace{.1pt}\discretionary{.}{%
}{.}\hspace{.4pt}571996}}}


\bibitem{crowbar}
{The New York Times}.
\newblock {SVG Crowbar}, 2017.
\newblock
  \href{https://nytimes.github.io/svg-crowbar/}{https://nytimes.github.io/svg-crowbar}.

\bibitem{tsandilas2020structgraphics}
\href{https://doi.org/10.1109/TVCG.2020.3030476}{T.~Tsandilas}.
\newblock \href{https://doi.org/10.1109/TVCG.2020.3030476}{{StructGraphics}:
  Flexible visualization design through data-agnostic and reusable graphical
  structures}.
\newblock \href{https://doi.org/10.1109/TVCG.2020.3030476}{{\em IEEE
  Transactions on Visualization and Computer Graphics (Proceedings of
  InfoVis)}}, \href{https://doi.org/10.1109/TVCG.2020.3030476}{27(2)},
  \href{https://doi.org/10.1109/TVCG.2020.3030476}{2021}.
  \href{https://doi.org/10.1109/TVCG.2020.3030476}
{DOI: {{%
10\hspace{.1pt}\discretionary{.}{%
}{.}\hspace{.4pt}1109\discretionary{/}{%
}{/}TVCG\hspace{.1pt}\discretionary{.}{%
}{.}\hspace{.4pt}2020\hspace{.1pt}\discretionary{.}{%
}{.}\hspace{.4pt}3030476}}}


\bibitem{p5js}
M.~Turner and L.~L. McCarthy.
\newblock {p5.js}, 2021.
\newblock \href{https://p5js.org/}{https://p5js.org}.

\bibitem{p5editor}
M.~Turner and L.~L. McCarthy.
\newblock {p5.js Web Editor}, 2021.
\newblock \href{https://editor.p5js.org/}{https://editor.p5js.org}.

\bibitem{Viegas_InfoVis_2004}
\href{https://doi.org/10.1109/INFVIS.2004.8}{F.~B. Vi{\'e}gas, E.~Perry,
  E.~Howe, and J.~Donath}.
\newblock \href{https://doi.org/10.1109/INFVIS.2004.8}{Artifacts of the
  presence era: Using information visualization to create an evocative
  souvenir}.
\newblock \href{https://doi.org/10.1109/INFVIS.2004.8}{In {\em IEEE Symposium
  on Information Visualization (InfoVis)}},
  \href{https://doi.org/10.1109/INFVIS.2004.8}{2004}.
  \href{https://doi.org/10.1109/INFVIS.2004.8}
{DOI: {{%
10\hspace{.1pt}\discretionary{.}{%
}{.}\hspace{.4pt}1109\discretionary{/}{%
}{/}INFVIS\hspace{.1pt}\discretionary{.}{%
}{.}\hspace{.4pt}2004\hspace{.1pt}\discretionary{.}{%
}{.}\hspace{.4pt}8}}}


\bibitem{ware2004information}
C.~Ware.
\newblock {\em Information Visualization: Perception for Design}.
\newblock Morgan Kaufmann Publishers Inc., 2004.

\bibitem{webb2008context}
\href{https://doi.org/10.1145/1385569.1385586}{A.~Webb and A.~Kerne}.
\newblock \href{https://doi.org/10.1145/1385569.1385586}{The in-context slider:
  a fluid interface component for visualization and adjustment of values while
  authoring}.
\newblock \href{https://doi.org/10.1145/1385569.1385586}{In {\em Proceedings of
  the ACM Conference on Advanced Visual Interfaces (AVI)}},
  \href{https://doi.org/10.1145/1385569.1385586}{2008}.
  \href{https://doi.org/10.1145/1385569.1385586}
{DOI: {{%
10\hspace{.1pt}\discretionary{.}{%
}{.}\hspace{.4pt}1145\discretionary{/}{%
}{/}1385569\hspace{.1pt}\discretionary{.}{%
}{.}\hspace{.4pt}1385586}}}


\bibitem{wickham2012glyph}
\href{https://doi.org/10.1002/env.2152}{H.~Wickham, H.~Hofmann, C.~Wickham, and
  D.~Cook}.
\newblock \href{https://doi.org/10.1002/env.2152}{Glyph-maps for visually
  exploring temporal patterns in climate data and models}.
\newblock \href{https://doi.org/10.1002/env.2152}{{\em Environmetrics}},
  \href{https://doi.org/10.1002/env.2152}{23(5)},
  \href{https://doi.org/10.1002/env.2152}{2012}.
  \href{https://doi.org/10.1002/env.2152}
{DOI: {{%
10\hspace{.1pt}\discretionary{.}{%
}{.}\hspace{.4pt}1002\discretionary{/}{%
}{/}env\hspace{.1pt}\discretionary{.}{%
}{.}\hspace{.4pt}2152}}}


\bibitem{wongsuphasawat2017voyager}
\href{https://doi.org/10.1145/3025453.3025768}{K.~Wongsuphasawat, Z.~Qu,
  D.~Moritz, R.~Chang, F.~Ouk, A.~Anand, J.~Mackinlay, B.~Howe, and J.~Heer}.
\newblock \href{https://doi.org/10.1145/3025453.3025768}{Voyager 2: Augmenting
  visual analysis with partial view specifications}.
\newblock \href{https://doi.org/10.1145/3025453.3025768}{In {\em Proceedings of
  the ACM Conference on Human Factors in Computing Systems (CHI)}},
  \href{https://doi.org/10.1145/3025453.3025768}{2017}.
  \href{https://doi.org/10.1145/3025453.3025768}
{DOI: {{%
10\hspace{.1pt}\discretionary{.}{%
}{.}\hspace{.4pt}1145\discretionary{/}{%
}{/}3025453\hspace{.1pt}\discretionary{.}{%
}{.}\hspace{.4pt}3025768}}}


\bibitem{wu2017}
S.~Wu.
\newblock {Film Flowers}.
\newblock In {\em Data Sketches}. CRC Press, 2021.
\newblock
  \href{https://shirleywu.studio/filmflowers/}{https://shirleywu.studio/filmflowers}.

\bibitem{wsj:redblueamerica}
R.~Yeip, S.~A. Thompson, and W.~Welch.
\newblock {A Field Guide to Red and Blue America}, 2016.
\newblock \textit{The Wall Street Journal}.
  \href{http://graphics.wsj.com/elections/2016/field-guide-red-blue-america/}{http://graphics.wsj.com/elections/2016/field-guide-red-blue-america}.

\bibitem{zhang2020dataquilt}
\href{https://doi.org/10.1145/3313831.3376172}{J.~E. Zhang, N.~Sultanum,
  A.~Bezerianos, and F.~Chevalier}.
\newblock \href{https://doi.org/10.1145/3313831.3376172}{{DataQuilt}:
  Extracting visual elements from images to craft pictorial visualizations}.
\newblock \href{https://doi.org/10.1145/3313831.3376172}{In {\em Proceedings of
  the ACM Conference on Human Factors in Computing Systems (CHI)}},
  \href{https://doi.org/10.1145/3313831.3376172}{2020}.
  \href{https://doi.org/10.1145/3313831.3376172}
{DOI: {{%
10\hspace{.1pt}\discretionary{.}{%
}{.}\hspace{.4pt}1145\discretionary{/}{%
}{/}3313831\hspace{.1pt}\discretionary{.}{%
}{.}\hspace{.4pt}3376172}}}


\end{thebibliography}
\end{document}